\documentclass[12pt,a4paper]{article}
\usepackage{amsmath}
\usepackage{amssymb}
\usepackage[sort]{cite}
\usepackage{hyperref}
\usepackage{graphicx}
\usepackage{color}
\usepackage{enumerate}
\usepackage{tikz}
\allowdisplaybreaks

\makeatletter
\@addtoreset{equation}{section}

\makeatother

\makeatletter
\let\old@makecaption=\@makecaption
\def\@makecaption{\small\old@makecaption}
\makeatother

\makeatletter
\let\old@startsection=\@startsection
\renewcommand{\@startsection}[6]{\old@startsection{#1}{#2}{#3}{#4}{#5}{#6\mathversion{bold}}}
\makeatother

\let\oldPhi=\Phi
\let\oldPsi=\Psi
\let\oldGamma=\Gamma
\let\oldSigma=\Sigma
\renewcommand{\Phi}{\mathnormal{\oldPhi}}
\renewcommand{\Psi}{\mathnormal{\oldPsi}}
\renewcommand{\Gamma}{\mathnormal{\oldGamma}}
\renewcommand{\Sigma}{\mathnormal{\oldSigma}}

\newcommand{\hypref}[2]{\ifx\href\asklfhas #2\else\href{#1}{#2}\fi}

\newcommand{\alg}[1]{\mathfrak{#1}}

\newcommand{\Tr}{\mathop{\mathrm{Tr}}}

\newenvironment{myeqnarray}{\arraycolsep0pt\begin{eqnarray}}{\end{eqnarray}\ignorespacesafterend}
\newenvironment{myeqnarray*}{\arraycolsep0pt\begin{eqnarray*}}{\end{eqnarray*}\ignorespacesafterend}

\def\[{\begin{equation}}
\def\]{\end{equation}}
\def\<{\begin{myeqnarray}}
\def\>{\end{myeqnarray}}

\ifx\href\asklfhas\newcommand{\href}[2]{#2}\fi

\begin{document}

\thispagestyle{empty}

\vspace{1cm}

\renewcommand{\thefootnote}{\fnsymbol{footnote}}
\setcounter{footnote}{0}

\begin{center}
{\Large\textbf{\mathversion{bold}Fermionic Correlators from Integrability}\par}

\vspace{1cm}

\textsc{Jo\~ao Caetano$^{a,b}$ and Thiago Fleury$^{b,c}$ }
\vspace{8mm}

\textit{$^{a}$ Laboratoire de Physique Th\'eorique de l' \'Ecole Normale Sup\'erieure,  \\
 PSL Research University,  CNRS,  Sorbonne Universit\'es, UPMC
Univ.\,Paris 06,  \\
 24 rue Lhomond, 75231 Paris Cedex 05, France}

\vspace{4mm}

\textit{$^{b}$ Perimeter Institute for Theoretical Physics, \\
Waterloo, Ontario N2L 2Y5, Canada}
\vspace{4mm}

\textit{$^{c}$
Instituto de F\'isica Te\'orica, UNESP - Univ. Estadual Paulista, \\
ICTP South American Institute for Fundamental Research, \\
Rua Dr. Bento Teobaldo Ferraz 271, 01140-070, S\~ao Paulo, SP, Brasil} 

\vspace{8mm}

\texttt{joao.caetano@lpt.ens.fr}\\
\texttt{tfleury@ift.unesp.br}\par\vspace{1cm}

\textbf{Abstract}\vspace{5mm}

\begin{minipage}{13.7cm}
We study three-point functions of single-trace 
operators in the 
$\mathfrak{su}(1|1)$
sector of planar $\mathcal{N}=4$ SYM borrowing several 
tools based on Integrability.
In the most general configuration of operators 
in this sector, we have found a determinant 
expression for 
the tree-level structure constants. We then compare 
the predictions 
of the recently proposed hexagon program against 
all available data. We have obtained a match  once additional 
sign factors are included when the two hexagon form-factors 
are assembled together to form the 
structure constants.
In the particular case of 
one BPS and two non-BPS operators
we managed to identify the relevant 
form-factors
with a domain wall partition function of a 
certain six-vertex model. This partition function can 
be explicitly evaluated and factorizes at all loops. 
In addition, we use this result to compute the 
structure constants
and show
that at strong coupling in the so-called BMN regime, its leading order contribution has a determinant expression.

\end{minipage}

\end{center}

\newpage
\setcounter{page}{1}
\renewcommand{\thefootnote}{\arabic{footnote}}
\setcounter{footnote}{0}

\tableofcontents
\section{Introduction}
Recently, 
a significant progress has been made in the computation of the 
structure constants of planar $\mathcal{N}=4$ 
SYM 
by integrability techniques. 
The use of integrability to tackle this 
problem was initiated 
mostly
in the papers \cite{Tseng,Roiban,TailoringI} 
and 
culminated in a
non-perturbative 
proposal formulated in 
\cite{hexagon}. 
This conjectured all-loop solution is 
grounded on a very 
stringy picture. 
The three-point functions are 
represented by a pair of pants corresponding to the well 
known 
idea of the splitting of a string into two other strings.
Upon cutting open this pair of pants one is left with two hexagons  patches
with their edges identified. 
The hexagons are then regarded as a sort of fundamental objects 
that inherit 
information about   
the initial and final 
states. 
In particular, in the integrability language 
the external states are characterized
by a set of parameters named Bethe rapidities and 
as a consequence
the hexagons form-factors 
are 
functions of these 
rapidities. 
Eventually in \cite{hexagon}  
it was possible to bootstrap completely the 
so-called hexagon form-factors
mostly by symmetry considerations.
Once they are known, 
the 
structure constants can be 
obtained 
by gluing a pair of these hexagons
and the final outcome is expressed in terms of sums over partitions of the Bethe rapidities of each operator.
Several checks of the hexagon program predictions 
against the perturbative data were already made in the original paper   
\cite{hexagon}. Additional 
checks were made 
both at strong and weak 
coupling in  
\cite{Eden3,Basso3,DiagonalI,DiagonalII,Kazama,ShotaLast}  
providing very strong support for the  
correctness of the hexagon  
solution to the structure constants problem.
 
In this paper,
we concentrate on operators sitting 
in the closed $\alg{su}(1|1)$ sector which is the
smallest sector containing fermionic excitations and consider their
asymptotic 
three-point functions. 
This means that we
take all the lengths 
involved to be large. 
In the hexagon language, the finite size corrections 
are controlled by the mirror particles and thus we can
safely
neglect them in this regime. 
One of the goals of this work is to check the predictions 
of the hexagon program for fermionic correlators 
against the perturbative data. 
We have found perfect agreement in all cases considered
provided 
we include some \textit{ad hoc}  
partition dependent additional signs in the hexagon program.
This rule differs from the original proposal of \cite{hexagon}, which already included some put-on signs to cohere with data, and
we do not 
have a convincing geometric explanation for
their origin. 

In the section \ref{general2}, 
we express the 
$\mathfrak{su}(1|1)$ primary operators
in terms of some {\it{polarization}} vectors
and directly compute 
the most general tree-level structure constant involving
three of these operators. 
We prove that the result 
admits a determinant expression  
depending on the  
Bethe rapidities parametrizing the excitations of the three
operators. 
In general, the result of a 
three-point function is a sum of many inequivalent 
tensor structures \cite{Sotkov1,Sotkov2,Spinning}. However
in all cases considered in this work there is only one 
tensor structure (and consequentially one structure
constant) and therefore it will be omitted everywhere.
We refer the reader to 
\cite{CaetanoFleury} for details.

In the section \ref{Secao3}, we 
apply the hexagon program for the $\alg{su}(1|1)$ sector.
Firstly, the case of 
one BPS and two non-BPS operators 
is considered.
We prove by deriving recursion relations
that the relevant hexagon form-factors for computing
the structure constants in those cases can be explicitly
evaluated and have 
a completely factorized form 
at all loops. 
Interestingly, the matrix part of
these hexagon form-factors can be viewed as a 
partition function of a certain six-vertex model
at any loop order. This fact is only true for operators in the   
$\alg{su}(1|1)$ sector. A similar setup but with operators 
in the $\alg{su}(2)$ sector was considered in the Appendix K of
\cite{hexagon} and the hexagon form-factors are domain wall partition
functions of a six-vertex model only at tree-level. This
is the expected result because at tree-level the 
three-point function
reduces to an off-shell scalar product 
\cite{su2offshell,su2offshell2,Wheeler}. In addition, 
we take the strong coupling limit of our results for the
structure constants
in the so-called BMN regime.
Surprisingly, we show that
the leading
contribution to the structure constants can be written
as a determinant for any number of excitations. 

The case of three non-BPS operators is also studied in the section
\ref{Secao3}. In \cite{CaetanoFleury}, the one-loop structure
constants for specific three $\alg{su}(1|1)$ operators were
computed both by finding the two-loop Bethe eigenstates and 
by evaluating all the relevant Feynman diagrams. We have checked 
numerically that
the hexagon program reproduces the 
results of 
\cite{CaetanoFleury}. 
The final answer for the structure constants 
in the hexagon program is given as a 
sum over partitions of three sets of Bethe rapidities 
(one for each operator) of the product of two hexagons
form-factors. 
It is clearly a quite demanding  task to explicitly compute 
them for a large number of excitations. 
It is very likely though 
that this solution can be further simplified at least
for some cases. 
Three instances where such simplifications were attained are the determinant 
expressions of section \ref{general2}
and subsection \ref{full3pt2nonbps},
the final expression for the structure
constants of three $\alg{su}(1|1)$ operators of \cite{CaetanoFleury}
and the results of \cite{ShotaLast} in the semiclassical limit.

\section{General tree-level structure constants 
in $\mathfrak{su}(1|1)$} \label{general2}

In this section, we consider the most general configuration 
of operators in the $\mathfrak{su}(1|1)$ 
sector
of $\mathcal{N}=4$ SYM. There are different embeddings 
of $\alg{su}(1|1)$ in the full superconformal group and they 
can be conveniently parametrized through some polarization 
vectors. 
This has a resemblance with the studies made in  the $\alg{su}(2)$ sector 
presented in \cite{KazamaPolarization}.  

The $R$-charge index contractions in the three-point 
functions 
considered here
are nicely accounted by the scalar products of the 
polarization vectors. 
It then remains to compute the 
dynamical part which is the most interesting one. 
At tree-level, we make full use of the fact that these operators are 
described by free fermions and we are able to derive 
a determinant 
expression for the structure constants. 

\subsection{Polarization vectors for $\alg{su}(1|1)$}

In order to parametrize the external operators in the three-point 
function, let us start by introducing a pair of 
\textit{polarization} vectors $Z_a$ and $W_a$, where $a=1,\ldots,4$  
are $\alg{su}(4)$ indices, satisfying the 
following normalization and orthogonality conditions
\begin{equation}
\bar{Z}^a Z_a =1 \, , \quad  \bar{W}^a W_a = 1 \, , \quad \bar{Z}^a W_a = 0 \, , \quad \bar{W}^a Z_a = 0 \, ,  
\end{equation}
with the bar standing for the 
usual complex conjugation $\bar{Z}^a \equiv (Z_{a})^*$. 

A state in the $\alg{su}(1|1)$ sector is built out of a scalar 
field $\Phi$ and a fermionic field $\Psi$ that 
we define in terms of the above polarization vectors by
\begin{equation}
\Phi = Z_a W_b\, \phi^{ab}  \, , \quad \Psi = W_a \,\psi^a - Z_c\, \psi^c \, ,   
\end{equation} 
where $\phi^{ab}$ and $\psi^c$ are the scalar and fermion fields
of $\mathcal{N}=4$ SYM and 
we have omitted the Lorentz index $\alpha$ of the fermions, 
as we fix it once and for all to take the value $\alpha=1$. 
We now want to show that the fields $\Phi$ and $\Psi$ 
form a representation of the tree-level algebra. 
For that let us define a supercharge  $\mathcal{Q}$ as follows
\begin{equation}
\mathcal{Q} \equiv \bar{Z}^a Q_a^1 + \bar{W}^a Q_a^1 \,,
\end{equation}
where $Q_a^\alpha$ is the standard bare 
supercharge that generates the usual supersymmetry transformations
on the fields
\begin{equation}
[ Q_ a^{\alpha} \, , \phi^{bc} ] = 
\delta^b_a \psi^{c \alpha} - \delta^c_a \psi^{b \alpha} \, , 
\quad \quad  [ Q_a^{\alpha} , \psi^b_{\beta}] = \delta^b_a 
F^{\alpha}_{\beta} \, ,
\end{equation} 
where $F^{\alpha}_{\beta}$ is the self-dual field-strength. 
We then observe that the relations  
$\mathcal{Q}\, \Phi = \Psi$ and $\mathcal{Q}\,\Psi = 0$ 
hold which imply a $\mathfrak{su}(1|1)$ representation.

A general $\alg{su}(1|1)$ primary operator can 
then be defined by 
specifying a pair of vectors $(Z_a,W_a)$.
For example, an operator $\mathcal{O}$ 
with $N$ excitations and length $L$ is defined by
\begin{equation} \label{oper}
\mathcal{O}(Z,W)=\sum_{1 \le n_1 <  \ldots 
< n_{N} \le L} \psi_{n_1,n_2,\ldots,n_N} \, \text{Tr} \,  
( \Phi \dots \underset{n_1}\Psi \dots 
 \underset{n_2}\Psi \dots \Phi ), 
\end{equation}
where the dependence in the polarization vectors is hidden in 
$\Phi$ and $\Psi$ and $\psi$ is a wave-function
(we are omitting the $\alg{su}(4)$ 
index $a$ to simplify the notation). 

The 
contraction 
of the $R$-charge indices 
between 
two given 
scalar fields 
parameterized by 
$(Z^{(1)},W^{(1)})$ and $(Z^{(2)},W^{(2)})$ respectively, 
gives the following contribution  
\begin{equation}
\begin{aligned}
\langle \Phi^{(1)} \Phi^{(2)} \rangle &= Z^{(1)}_{a}W^{(1)}_{b} Z^{(2)}_{c}W^{(2)}_{d} \, \langle \phi^{ab} \phi^{cd} \rangle \\
&= \det [ \{ Z^{(1)}, W^{(1)}, Z^{(2)}, W^{(2)} \} ]\\
&\equiv \langle 12\rangle \, . 
\end{aligned}
\end{equation}
Analogously, we have that the contraction of a scalar $\Phi^{(1)}$ and a conjugate scalar $\bar\Phi^{(2)}$ is given by
\begin{equation}
\begin{aligned}
\langle \Phi^{(1)} \bar\Phi^{(2)} \rangle &= Z^{(1)}_{a}W^{(1)}_{b} \bar{Z}^{(2)\, c} \,\bar{W}^{(2) \, d} \, \langle \phi^{ab} \phi_{cd} \rangle \\
&=Z^{(1)}_{a}W^{(1)}_{b} \left( \bar{Z}^{(2)\, a} \,\bar{W}^{(2) \, b} - \bar{Z}^{(2)\, b} \,\bar{W}^{(2) \, a} \right) \\
&\equiv \langle 1 \bar{2}\rangle \, . 
\end{aligned}
\end{equation}
Finally for the fermions, one has
\begin{equation}
\begin{aligned}
\langle \Psi^{(1)} \bar\Psi^{(2)} \rangle &= (Z^{(1)}_{a}-W^{(1)}_{a})  (\bar{Z}^{(2)\,b}-\bar{W}^{(2)\,b}) \, \langle \psi^{a} \bar\psi_{b} \rangle\\
&=  (Z^{(1)}_{a}-W^{(1)}_{a})  (\bar{Z}^{(2)\,a}-\bar{W}^{(2)\,a})\\\,
&\equiv [ 1 \bar{2} ]\,.
\end{aligned}
\end{equation}

The setup we will be considering in this section
is formed by three operators of the type (\ref{oper}), 
each one characterized by a pair of polarization vectors 
$(Z^{(i)},W^{(i)})$ for $i=1,2,3.$ Moreover, in order 
to have a non-zero structure constant, we conventionally 
take the operator $\mathcal{O}_2$ to have the 
antichiral fermions, that is  
\begin{equation} \label{oper1}
\mathcal{O}_2(\bar{Z}^{(2)},\bar{W}^{(2)})=
\sum_{1 \le n_1 <  \ldots 
< n_{N_2} \le L_2} \psi^{(2)}_{n_1,n_2,\ldots,n_{N_2}} \, \text{Tr} \,  
( \bar\Phi^{(2)} \dots \underset{n_1}{\bar\Psi}^{(2)} \dots 
 \underset{n_2}{\bar\Psi}^{(2)} \dots \bar\Phi^{(2)} ) \, .
\end{equation}
We will now make use of this parametrization of 
the operators to compute the tree-level structure constants.

\subsection{Tree-level three-point functions as a determinant} 
\label{sec:Thedeterminant}

At tree-level, the wave-function $\psi^{(i)}$
associated to 
the operator $\mathcal{O}_i$ is 
given by the standard Bethe wave-function for a 
free fermion system that 
follows from
the requirement 
that it diagonalizes the one-loop 
$\alg{su}(1|1)$ Hamiltonian\footnote{It is simple to use the one-loop perturbative
results of the Appendix B of 
\cite{CaetanoFleury} to show that the one-loop
dilatation operator  acting on operators 
built out of $\Phi$ and $\Psi$  for a general polarization vector reduces to the usual 
$\mathfrak{su}(1|1)$ Hamiltonian, i.e. it is proportional
to the difference of the identity and the superpermutator
as expected. }
(more details can be found in \cite{CaetanoFleury}).
It is given by
\begin{eqnarray} \label{wavefunction}
\psi^{(i)}_{n_1, n_2, \ldots, n_{N_i}} = \sum_{P} \, 
(-1)^P
\, {\rm{exp}}(i p^{(i)}_{\sigma_P (1)} n_1 + 
i p^{(i)}_{\sigma_P (2)} n_2 + \ldots + i 
p^{(i)}_{\sigma_P (N_{i})} n_{N_i}) \, ,
\end{eqnarray}
where $P$ indicates sum over all possible permutations 
$\sigma_P$ of the elements $\{1, . . . , N_i \}$, 
and 
$(-1)^P$
is the sign of the permutation.
In addition, the momenta satisfy the Bethe equations
\begin{equation}
e^{i p^{(i)}_j L_{i}} = 1 \, .  
\end{equation}

\begin{figure}[t!]
\centering
\includegraphics[width=90mm]{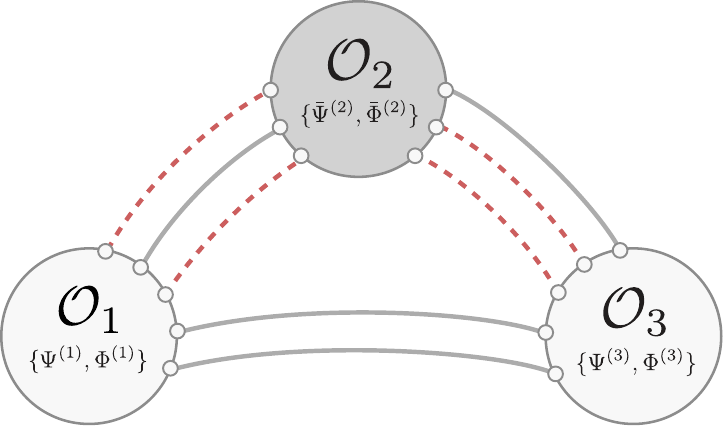}
\caption{This figure illustrates the 
Wick contractions for the computation of  
the tree-level 
three-point function in the most general 
$\alg{su}(1|1)$ setup. The dashed (solid) lines correspond to 
fermions (scalars) propagators. 
It is clear from the figure that one should 
multiply the three  
wave-functions corresponding to each of the 
operators and perform a sum over the positions of 
the fermionic excitations. In our conventions, all the
operators are oriented clockwise. 
\label{treelevel}}
\end{figure}
The tree-level structure constant is 
simply given by the product 
of the three wave-functions with the positions of the excitations 
of each operator summed over, see figure \ref{treelevel} for clarity. 
Concretely, we have the following nested sums to evaluate
\begin{equation}\label{eq:brute}
C_{123} =  \mathcal{R}
\hspace{-5mm} \sum_{\substack{1\le n_1 < \ldots 
< n_{N_1} \le l_{12}  \\
1 \le m_1 < \ldots < m_{N_3} \le l_{23}}}   
\psi^{(1)}_{L_1 - n_{N_1} +1,\ldots ,L_1 -n_{1}+1}  
\psi^{(2)}_{n_1, \ldots ,n_{N_1}, l_{12}+ m_1, \ldots, l_{12}+m_{N_3}} 
\psi^{(3)}_{l_{23}-m_{N_3}+1, \ldots ,l_{23}-m_{1}+1} \,, 
\end{equation}
where $\mathcal{R}$ includes the contribution from the $R$-charge 
contractions and the normalization factor and reads
\begin{equation}
 \mathcal{R} = \sqrt{ \frac{ L_1 L_2 L_3}{{\mathcal{N}}^{(1)} 
 {\mathcal{N}}^{(2)} {\mathcal{N}}^{(3)}}}  \,\,\, \langle 13 
 \rangle^{l_{13}} [ 3\bar{2}]^{N_{3}} \langle 3\bar{2}\rangle ^{l_{23} - N_{3}} 
 [ 1\bar{2} ]^{N_{1}} \langle 1\bar{2} \rangle^{l_{12} - N_{1}}\,,
\end{equation}
where $\mathcal{N}^{(i)}$ is 
the norm of the wave-function $\psi^{(i)}$ and $l_{ij}$ is 
the number of contractions between operators $i$ and $j$.

Given that the wave-functions in (\ref{wavefunction}) are 
completely antisymmetric in all their arguments, we can extend 
the sums in (\ref{eq:brute}) at the price of introducing a trivial 
overall combinatorial factor. Plugging their explicit expressions, 
we are left with 
\begin{eqnarray}
\frac{C_{123}}{\mathcal{R}}=  \sum_{\{n_i\},\{m_i\}} 
\sum_{P,\,Q,\,S} \frac{(-1)^{P+Q+S}}{N_1! N_3!}\,\times 
 \, \hspace{65mm} \\
\prod_{a=1}^{N_1} \prod_{b=1}^{N_3}
e^{i p^{(1)}_{P(N_1 - a +1)} (1 - n_a)} 
e^{i p^{(2)}_{S(a)} n_a} e^{i p^{(2)}_{S(N_1 + b)} (l_{12} + m_b)}
e^{i p^{(3)}_{Q(N_3 - b +1)} (l_{23}-m_b + 1)} 
 \, ,  \hspace{-15mm} \nonumber
\end{eqnarray} 
where we have simplified the wave-function of the operator 
$\mathcal{O}_1$ 
by using the Bethe equations. Note that the sums over 
$n_i$ and $m_i$ are not ordered anymore 
and run through the full range  $1\leq n_i \leq l_{12}$ 
and $1 \leq m_i\leq l_{23}$. 
It is now simple to perform the sums over 
$n_i$ and $m_i$ as they are  
geometric series. This results in 
\[
\frac{C_{123}}{\mathcal{R}}= 
\sum_{P,\,Q,\,S} 
\frac{(-1)^{P+Q+S}}{N_1!N_3!}\,\prod_{a=1}^{N_1} \prod_{b=1}^{N_3}\,
\frac{1- e^{i \left(p^{(2)}_{S(a)}-p^{(1)}_{P(N_1+1-a)}\right)l_{12}}}
{e^{-ip^{(2)}_{S(a)}} -e^{-ip^{(1)}_{P(N_1+1-a)}} } \times
\frac{e^{-i \left(p^{(2)}_{S(N_1+b)}-p^{(3)}_{Q(N_3+1-b)}\right)l_{23}}-1}
{e^{-ip^{(2)}_{S(N_1 + b)}} -e^{-ip^{(3)}_{Q(N_3+1-b)}} }.
\]
It is not hard to recognize this expression as being,
apart for some signs, the definition 
of the determinant of a 
$N_2$ by $N_2$ matrix formed by two blocks namely,
\[ \label{determinantI}
C_{123}=\mathcal{R}\, 
(-1)^{\frac{N_1 (N_1 -1)}{2}} \,
(-1)^{\frac{N_3 (N_3 -1)}{2}} \, 
\det_{1 \leq j,k \leq N_2} \left[C^{(1)}_{jk} 
\oplus C^{(3)}_{jk}    \right]
\]
with the blocks being
\[
\begin{aligned} \label{determinant}
&C^{(1)}_{jk} = 
\frac{1-e^{i \left(p^{(2)}_j - p^{(1)}_k \right) 
l_{1 2}} }
{e^{-i p^{(2)}_j} -e^{-i p^{(1)}_k}}\,, \,\,\,\,\, j=1,\dots, N_2,\,\, k=1,\dots,N_1\,,\\
& C^{(3)}_{jk} = 
\frac{e^{-i \left(p^{(2)}_j - p^{(3)}_k \right) l_{23}}-1}
{e^{-i p^{(2)}_j} -e^{-i p^{(3)}_k}}\,, \,\,\,\,\, 
j=1,\dots, N_2,\,\, k=1,\dots,N_3\,.
\end{aligned}
\]
This is the main result of this section. In 
what follows we will consider a few limits of this expression.
\paragraph{Extremal limit}
In the extremal limit $L_2=L_1+L_3$ which implies that 
$l_{23}=L_3$ and $l_{12}=L_1$. Inserting these conditions 
on the previous formula, it gets simplified once we use 
the Bethe equations and both blocks get a similar form 
\[
C_{jk}^{(1)} =  
\frac{1-e^{i p^{(2)}_j  L_1}}{e^{-i p^{(2)}_j} -e^{-i p^{(1)}_k}} \,, 
\quad \quad \quad 
C_{jk}^{(3)} = - 
\frac{1-e^{i p^{(2)}_j  L_1}}{e^{-i p^{(2)}_j} -e^{-i p^{(3)}_k}} \, .
\]
It immediately follows that this is a Cauchy 
matrix and one can use the Cauchy determinant formula
to obtain
\[
C_{123} = \mathcal{R} \, (-1)^{N_3} \, \prod\limits_{i=1}^{N_2}
{\left(1- e^{i p^{(2)}_i L_1}\right)} 
\frac{\prod\limits_{k=1}^{3}\prod\limits_{i>j}^{N_k} 
\alg{f}^{(kk)}_{ij}\prod\limits_{i=1}^{N_1}\prod\limits_{j=1}^{N_3} 
\alg{f}^{(13)}_{ij}}
{\prod\limits_{i=1}^{N_2}
\prod\limits_{j=1}^{N_1} \alg{f}^{(21)}_{ij}
\prod\limits_{i=1}^{N_2}\prod\limits_{j=1}^{N_3} \alg{f}^{(23)}_{ij}}\,,
\]
where we have defined
\[ \label{deff}
\alg{f}_{ij}^{(km)} \equiv e^{-ip^{(k)}_i}-e^{-ip^{(m)}_j}\,.
\]
In addition to this, 
the extremal case gets modified by the contribution 
coming from the one-loop mixing of the single-trace  
$\mathcal{O}_2$ with the double-trace operators. 
The calculation of this extra piece is outside the scope 
of the present paper and we leave it for future work.
\paragraph{Reduction to the formula of \cite{CaetanoFleury}}
Another limit where the determinant (\ref{determinant}) 
gets factorized is the configuration 
considered in \cite{CaetanoFleury}. In that  setup, one 
sets $l_{23}=N_3$ which leads to
\[
|C_{123}| = \mathcal{R} \left|  \, \prod\limits_{i=1}^{N_1}
{\left(1- e^{i p^{(1)}_i L_2}\right)} \frac{\prod\limits_{k=1}^{3}
\prod\limits_{j<i}^{N_k} \alg{f}^{(kk)}_{ji}}
{\prod\limits_{i=1}^{N_2}\prod\limits_{j=1}^{N_1} 
\alg{f}^{(21)}_{ij}} \, \right| \,.
\]
where $\alg{f}_{ij}^{(km)}$ was defined in (\ref{deff}).

\section{Hexagon program for fermionic correlators}\label{Secao3}

In this section, we will compute three-point functions of operators
containing fermionic excitations using the hexagon program of 
\cite{hexagon}. This method generates all-loop 
predictions
for the structure constants which as 
we will see match the results of the previous section 
when expanded at leading order. Firstly, we will briefly review 
the definition of the 
hexagon form-factor. We then show that the relevant hexagon 
for the three-point function 
of 
one BPS and two non-BPS
operators in the $\mathfrak{su}(1|1)$ sector 
has the interpretation of a domain wall partition function 
of a certain six-vertex model. 
We further prove that it has a completely 
factorized form to all loops. 
We perform checks with the available data for 
fermionic correlators and point out the need of some 
additional relative signs 
when the two hexagon form-factors are combined together
to form the three-point function  in order to get a match.

\subsection{Fermionic hexagons}
\begin{figure}[t!]
\centering
\includegraphics[width=70mm]{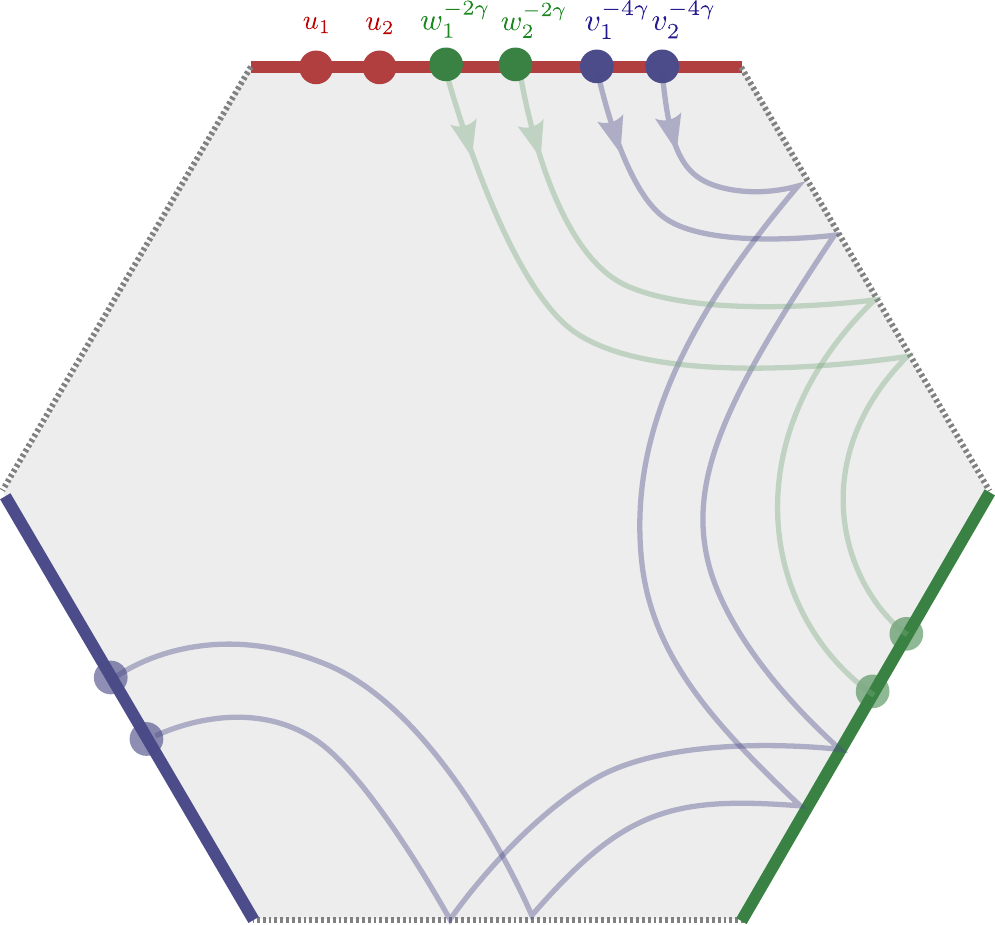}
\caption{When computing three-point 
functions using the hexagon program, we only need 
to consider the hexagon form-factor with excitations 
in a single edge, say the red edge in the figure. When 
more than one operator is excited, some of the excitations 
on the red edge will be mirror transformed. In the figure 
we identify the rapidities of the corresponding mirror 
transformed excitations by the upper symbol $\gamma$. 
This corresponds to having
moved them from the red edge to other edges 
of the hexagon. In particular an even number of such mirror transformations move them to other physical edges, represented in green and blue in the figure. The particular edge where they end up depends on the sign and number of mirror transformations applied to them. The conventions we use here are illustrated in the figure.
\label{crossings}}
\end{figure}

The fundamental excitations of an operator in the hexagon 
program
transform in the bifundamental representation of a centrally extended  
$\mathfrak{su}(2|2) \times \mathfrak{su}(2|2)$ algebra. They 
are labeled by two 
indices $(A \, , 
 \dot{A})$. In our conventions, these indices take the 
values 1 to 4 with $a=1,2$ being bosonic indices and $\alpha=3,4$ 
fermionic ones. Throughout this section we will be 
considering fermionic excitations which carry both
one bosonic and one fermionic index.

The hexagon form-factor in the string frame 
with $N$ excitations with rapidities $u_i$ 
in one physical  
edge (see figure \ref{crossings}) of the hexagon is given by 
\cite{hexagon} 
\begin{eqnarray}
h^{A_1 \dot{B}_1,\ldots,A_N \dot{B}_{N}}_{\rm{string}}(u_1,\ldots,u_N)=
(-1)^{f} \prod_{i < j} h(u_i, u_j) \langle \chi_N^{\dot{B}_N}  
\ldots \chi_{1}^{\dot{B}_{1}}| \, 
\mathcal{S} 
\,| 
\chi_{1}^{{A}_{1}} \ldots  \chi_{N}^{{A}_{N}} \rangle \, ,      
\label{eq:hexagonform}
\end{eqnarray}
where  $f = \sum_{i < j} {\rm{grad}}(\dot{B}_i) \,  
{\rm{grad}} (A_j) $ and grad means the grading of the corresponding 
index being equal to zero for bosonic indices and
to one for fermionic indices.
In the formula above, 
$\mathcal{S}$ 
is the $\mathfrak{su}(2|2)$ $S$-matrix in 
the string frame 
\cite{StringSpin,Foundations,ReviewIntegrability}
with the overall multiplicative
constant set to one and
$(\chi^A, \chi^{\dot{A}})$ are states
in the fundamental of 
$\mathfrak{su}(2|2) \times \mathfrak{su}(2|2)$. In Appendix \ref{Smatrix} we present the 
explicit form of the $S$-matrix used here. 
In order to evaluate the matrix part of the hexagon form-factor, 
one uses 
\begin{equation}
\langle \chi_N^{\dot{B}_N}  
\ldots \chi_{1}^{\dot{B}_{1}}| \,  
\chi_{N}^{{A}_{N}} \ldots  \chi_{1}^{{A}_{1}} \rangle = h^{A_1 \dot{B}_1} 
\ldots \, h^{A_N \dot{B}_N} \, , \label{eq:rulescontractingI}  
\end{equation}
and the only nonvanishing components are
\begin{equation} 
h^{1 \dot{2}}=-h^{2\dot{1}}=1 \, ,  \quad \quad 
h^{3 \dot{4}}=-h^{4\dot{3}}=-i \, . \label{eq:rulescontractingII}  
\end{equation}
Finally, the function $h(u,v)$ is defined by
\begin{equation}
h(u,v)= \frac{x_u^- - x_v^-}{x_u^- - x_v^+} \frac{1 - 1/x_u^- x_v^+}{1-1/x_u^+ x_v^+} \frac{1}{\sigma(u,v)} \, . 
\label{eq:hpequeno} 
\end{equation}
The variables $x$ are Zhukowsky variables satisfying
$x_u+1/x_u=u/g$ with $g$ the coupling constant and 
$x_u^{\pm}=x\bigl(u \pm \frac{i}{2} \bigr)$. Moreover, $\sigma(u,v)$ is half 
the dressing phase of \cite{DressingPhase}.

When computing a three-point function, we first transfer all 
the excitations of the three operators to one of the physical 
edges of the hexagon, see figure \ref{crossings}. This is done 
by performing successive mirror transformations on the excitations 
of certain operators. 
These mirror transformations correspond to an 
analytic 
continuation of the hexagon  in the rapidities of the 
corresponding excitations. In Appendix \ref{Smatrix} we present 
the transformation that the hexagon form-factor undergoes 
by
this analytic continuation for the fermionic excitations.  In addition 
to this, we will compare the predictions for the structure 
constants obtained using the hexagon program with the available 
weak coupling data. In order to perform these checks, we first 
do the computations using the string frame where the mirror 
transformations are implemented in a simple manner and 
then map the result to the spin-chain frame
\cite{BeisertSmatrix,BeisertSmatrixII}. The two frames are 
related by a phase depending on the momenta of the excitations 
and that is described in Appendix \ref{Smatrix}.

\subsection{All-loop factorization for 
1 BPS and 2 non-BPS operators}

In this subsection, we compute the structure constants
of two non-BPS fermionic operators and one BPS operator using the hexagon
program. We will further provide a closed expression for the hexagon form-factors in a completely factorized form. 

\begin{figure}[t!]
\centering
\includegraphics[width=150mm]{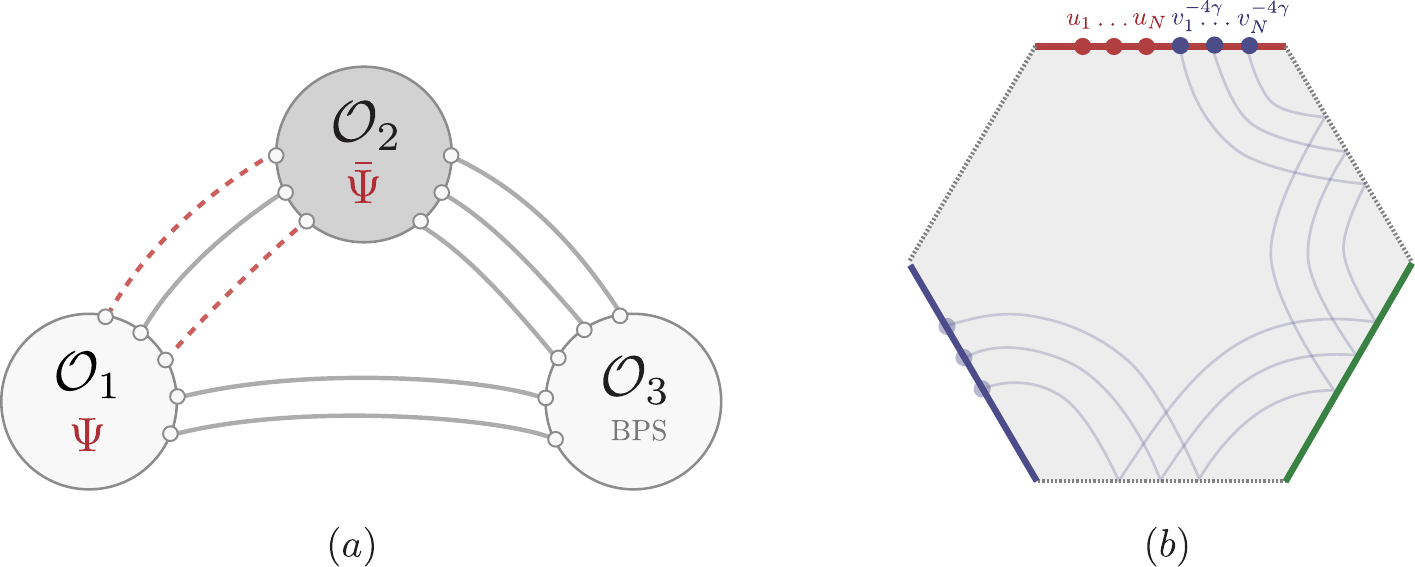}
\caption{(a) We represent the tree-level 
configuration used in this section. The operator 
$\mathcal{O}_1$ ($\mathcal{O}_2$) contains 
$\Psi$ ($\bar{\Psi}$) 
excitations,  $\mathcal{O}_3$ is BPS and it has no excitations. 
(b) The hexagon 
form-factor used in this section contains two types 
of excitations associated to the two non-BPS operators. A set 
of those excitations ($v_i$) are mirror transformed four 
times to the right (hence the minus sign in $-4\gamma$) and 
end up in the physical edge belonging to 
the operator $\mathcal{O}_1$. Alternatively we 
could have started with the excitations in the top edge 
with the positions of $u_i$ and $v_i$ reversed and 
performed two mirror transformations on the $v_i$ to the 
left. The 
first option is used in 
the main text as it makes 
the matrix part simpler.
\label{tree3}}
\end{figure}

The non-BPS operators are in the $\mathfrak{su}(1|1)$ sector
and 
contain a single type
of fermionic excitations.
To fix the conventions, we choose a setup of three operators 
in which the operator $\mathcal{O}_1$ has the excitations
$\Psi \equiv \chi^{ 3 \dot{1}}$, the operator 
$\mathcal{O}_2$ has
the excitations $\bar\Psi \equiv \chi^{2 \dot{4}}$ and the 
operator
$\mathcal{O}_3$ is BPS, see figure \ref{tree3}(a).

In order to use the defining expression 
for the hexagon
form-factors of (\ref{eq:hexagonform}), we need to move all
the excitations to the upper edge of the hexagon. There are two possible
ways of moving the excitation in the second physical edge to the upper
edge. One can perform either one crossing transformation\footnote{A crossing transformation corresponds to two mirror transformations. We denote a mirror transformation of a function $f(u)$ by $f(u^{\gamma})$.} or minus two
crossing transformations. We will choose the second possibility, see figure \ref{tree3}(b), for reasons of simplicity as will become clear below. Under this double crossing transformation, the fermionic excitations get their sign flipped according to the formula (\ref{crossrules}) of Appendix \ref{Smatrix}. Taking these signs into account, we
get that the central object of the three-point function 
for this setup is the following hexagon form-factor which we 
denote by $h_{\Psi_1 \ldots \Psi_N | \bar\Psi_1 \ldots 
\bar\Psi_N}$ and 
reads
\begin{align} 
 h_{\Psi_1 \ldots \Psi_N | \bar\Psi_1 \ldots \bar\Psi_N} 
 \equiv 
(-1)^N \prod_{i < j} h(u_i, u_j)\,
h(v_i^{-4\gamma}, v_j^{-4\gamma})
\prod_{i , j} h(u_i, v_j^{-4\gamma})  
\times [ \texttt{matrix part} ]
\,, \nonumber
\\[0.5em]
[ \texttt{matrix part} ] =  
\langle \chi_{v^{-4\gamma}_N}^{\dot{4}}  
\dots \chi_{v^{-4\gamma}_1}^{\dot{4}}  \chi_{u_N}^{\dot{1}} 
\dots   \chi_{u_1}^{\dot{1}}|\, \mathcal{S} \,| 
\chi_{u_1}^{{3}} \ldots \chi_{u_N}^{{3}} 
\chi_{v_1^{-4\gamma}}^{{2}} \ldots 
\chi_{v_N^{-4\gamma}}^{{2}}\rangle\,.
\hspace{11mm}  \label{hpsipsi}
\end{align}

Recall that we evaluate this form-factor in the string frame 
normalization. When we will later compare with data, we will then map it to the spin-chain frame using the conversion factor in formula (\ref{conversion}).
As 
an illustration, let us first compute the simplest case namely
$h_{\Psi | \bar\Psi}(u,v)$. 
We find    
\begin{eqnarray}
h_{\Psi | \bar\Psi}(u,v) =
-h(u,v^{-4 \gamma})  \langle \chi_{v^{-4 \gamma}}^{\dot{4}} \,   
\chi_{u}^{\dot{1}}| \, \mathcal{S}\, | 
\chi_{u}^{{3}} \, \chi_{v^{-4 \gamma}}^{{2}} \rangle = 
- \frac{i}{h(v,u)} K_{u v^{-4 \gamma}}  \, , 
\end{eqnarray}
where\footnote{Similarly 
we have $h_{\bar\Psi|\Psi} (u,v) = -h_{\Psi|\bar\Psi} (u,v)$.} the $S$-matrix element $K$ is defined in the Appendix A. Moreover,
we have used the following
properties
of the dressing phase to write $h(u,v^{-4 \gamma})$ in terms 
of $h(v,u)$,
\begin{equation}
\sigma (u^{2 \gamma}, v) \sigma(u,v) = \frac{(1 - 1/x^+_u x^+_v)}{(1 - 
x^-_u/ x^-_v)}
\frac{(1- x^-_u/x^+_v)}{(1 - 1/x^+_u x^-_v)} \, , \quad \quad \sigma(u,v)=
\frac{1}{\sigma(v,u)} \, .  
\end{equation}

Let us consider now the general case when there are $N$ 
excitations 
$\Psi$
in the upper edge of the hexagon and $N$ excitations 
$\bar\Psi$ in the second physical
edge of the hexagon. 
Using the formulae (\ref{crossx}), we can immediately 
write the pre-factor in front of the matrix part 
in expression (\ref{hpsipsi}) after the inverse 
crossing transformation of the set of rapidities $\{v\}$, to get
\begin{equation}
h_{\Psi_1 \ldots \Psi_N | \bar\Psi_1 \ldots \bar\Psi_N}
= 
(-1)^{N}  \frac{\prod_{i < j}^N h(u_i,u_j)
\prod_{i < j}^N h(v_i,v_j)}{ \prod_{i,j}^N h(v_i,u_j)}\times[ \texttt{matrix part} ] \, .
\label{eq:thehexagondyma}
\end{equation}

One way of evaluating the matrix part above is to first scatter 
the excitations $\chi^{3}_{u_i}$ among themselves to put them 
in descending 
order. We can then scatter them with all 
the other $\chi^{2}_{{v_i}^{-4 \gamma}}$. According 
to (\ref{saction}) this scattering will in general produce 
several terms where the 
indices can either be conserved or get swapped. Due to the one particle 
form-factor (\ref{eq:rulescontractingII}), the only non-zero $S$-matrix element occurs for the case where all the excitations $\chi^{3}_{u_i}$ swap their indices with 
$\chi^{2}_{{v_i}^{-4 \gamma}}$. Finally, we
scatter the resulting $\chi^{3}_{{v_i}^{-4 \gamma}}$ to put them 
in descending order as 
well. 
The first and last step of this procedure where excitations of the same species scatter among themselves results in a trivial factor as the only $S$-matrix element playing a role is
$D_{ij}=-1$.  

Interestingly, the non-trivial part of this scattering 
process turns out to be equivalent up to a phase
to the computation of a partition function that
resembles a domain wall partition function of a certain six-vertex model
as illustrated in figure \ref{fig:domainwall}(a) with the six
nonzero vertices of figure \ref{fig:sixvertexmodel}.  

The fact that we have a six-vertex model at any value of the 
coupling is remarkable and it is not true in general for other sectors. 
In the Appendix K of \cite{hexagon} for example, two non-BPS
$\mathfrak{su}(2)$ operators were considered and they only have
a six-vertex model at tree-level.  

\subsubsection{Factorization of the domain wall 
partition function}

\begin{figure}[t!]
\centering
\includegraphics[width=130mm]{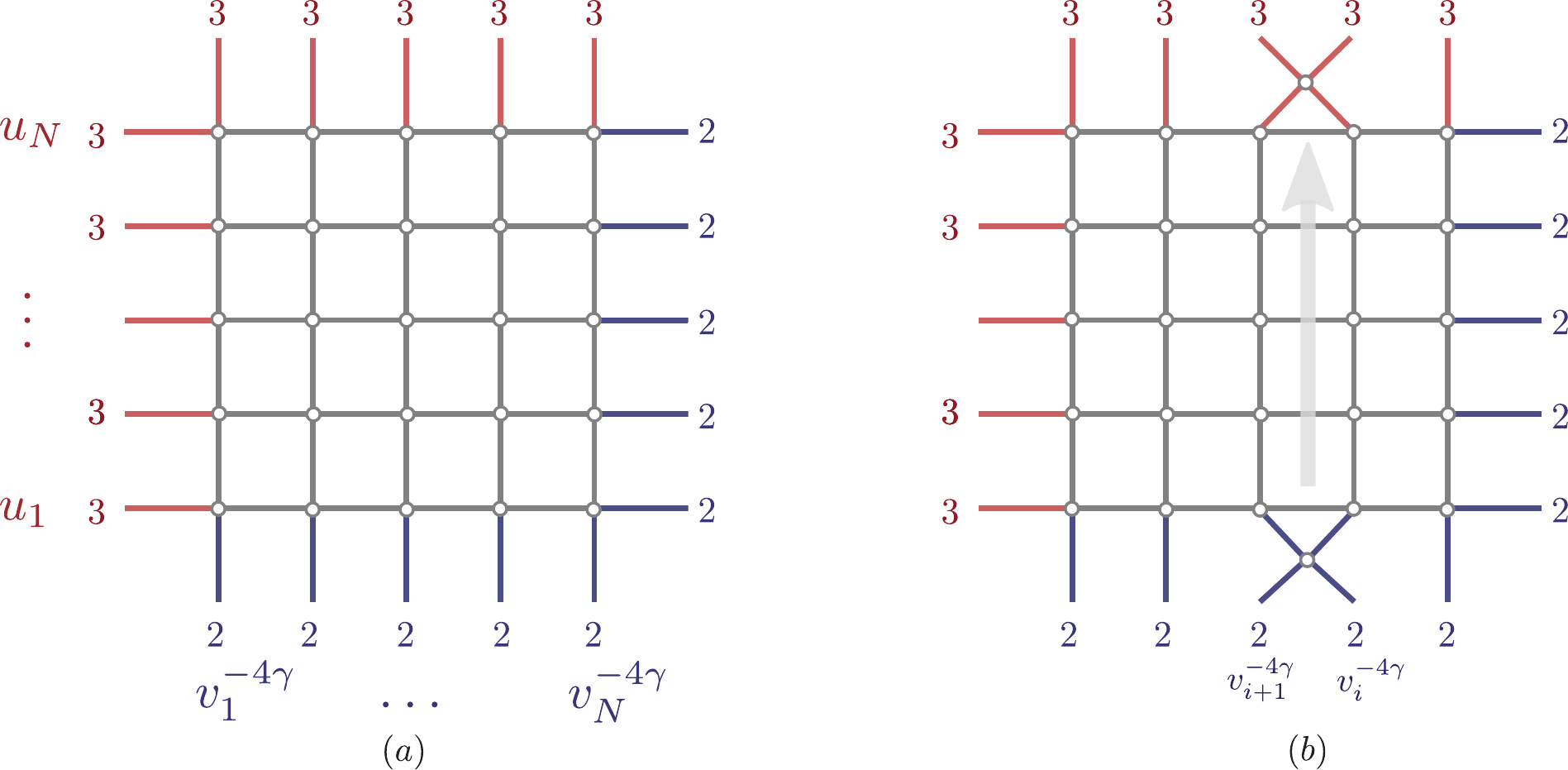}
\caption{(a) The
partition function appearing in the computation
of the matrix part of the hexagon form-factor. The  
vertices 
at any value of
the coupling are given in the figure \ref{fig:sixvertexmodel}. 
(b) The procedure to prove the relation \ref{second} of 
$\mathcal{P}_N$ goes as follows. Inserting an 
additional vertex to the bottom of the partition function grid,
it is possible to move it to the top using the 
Yang-Baxter equation with some rapidities crossed several times 
and remove it. The result is proportional to a grid 
with the two columns swapped. A similar procedure can be used 
to prove the relation \ref{third} of $\mathcal{P}_N$ but this time 
the procedure involves two neighboring lines. 
\label{fig:domainwall}}
\end{figure}

In this subsection, we will derive a closed expression
for the partition function of 
Figure \ref{fig:domainwall} valid 
at any value of the coupling constant.
We will denote the 
partition function by
${\rm{\mathcal{P}}}_N\left(\{u_1,\dots,u_N\},\{v_1, \dots, v_N\}\right)$. 
From the properties of the $S$-matrix, we can 
immediately infer the following relations

\begin{enumerate}[I.]
 \item 
 ${\mathcal{P}}_1(\{u_1\},\{v_1\}) = K_{u_1 {{v_1}^{-4 \gamma}}} \, ,$
\vspace{2mm} \label{first}
\item
${\mathcal{P}}_N(\{ u \},\{ v_1, \ldots ,v_i, v_{i+1}, \ldots, v_N \})
=- A_{v_i v_{i+1}}  \label{second} \, 
{\mathcal{P}}_N(\{ u  \},\{ v_1, \ldots, v_{i+1} , v_i \ldots , v_N \}) \, , $
\vspace{2mm}
\item
${\mathcal{P}}_N(\{ u_1, \ldots, u_i , u_{i+1}, \ldots, u_N \},\{ v \})
=- A_{u_i u_{i+1}}  \, 
{\mathcal{P}}_N(\{ u_1 , \ldots , u_{i+1}, u_i, \ldots, u_N   \},\{ v \}) \, . \label{third}$ 
\end{enumerate}

\begin{figure}[t!]
\centering
\includegraphics[width=155mm]{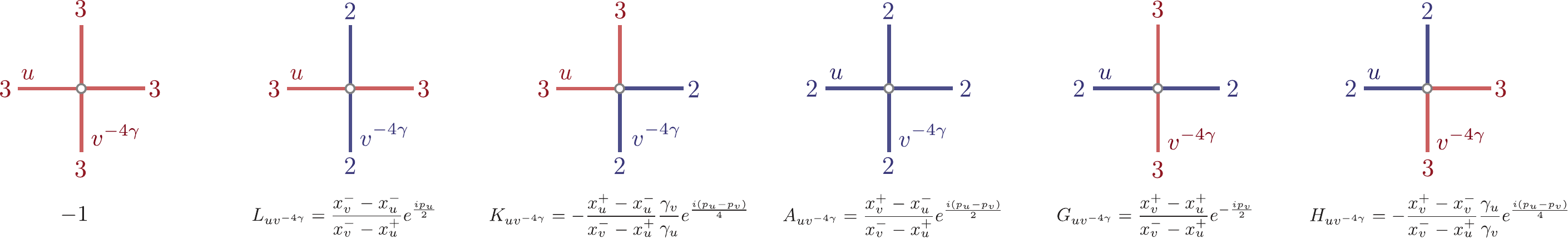}
\caption{The nonzero six vertices used for computing the 
partition function and their respective
weights which are equal to the components of
the string frame $S$-matrix, see Appendix \ref{Smatrix}.
\label{fig:sixvertexmodel}}
\end{figure}

The  relation \ref{first} simply follows from the fact that 
for $N=1$ the partition function reduces to the weight of 
the third vertex
given in figure \ref{fig:sixvertexmodel}. The relations \ref{second} and \ref{third} follow from using repeatedly the 
Yang-Baxter equation\footnote{In order to apply Yang-Baxter 
to this case, one should apply 
crossing transformations to some of the rapidites 
because the variables $v_i$ appear as 
$v_i^{-4 \gamma}$.} and the vertices 
in figure \ref{fig:sixvertexmodel} as illustrated 
in figure \ref{fig:domainwall}(b) (see also 
\cite{Garbali,Foda}).

As a first step to compute this partition function, we can use 
the previous properties to immediately infer its dependence on 
the phases of momenta $e^{i p_{u_k}}$, $e^{ip_{v_k}}$ and 
$\gamma_{u_k}$, $\gamma_{v_k}$ where 
$\gamma_{u} = \sqrt{i(x^-_u - x^+_u)}$.
Consider the top horizontal line. It is not difficult to see 
that the only allowed vertices on that line are the first, 
third and fifth of
figure \ref{fig:sixvertexmodel}. Moreover, the third vertex always appear only one time for each configuration on that line. Once it is used, then the whole line gets frozen. This vertex is the only among the allowed ones for the top line that depends on the momenta $p_{u_N}$ and $\gamma_{u_N}$. Therefore we can determine that the dependence of the whole partition function on these quantities comes from the weight of the third vertex.
 
 A similar analysis can be performed on the first vertical line. The only allowed vertices are the first, second and the third ones and again the third vertex necessarily appears only one time in every configuration on that line. Analogously, given the weights of these three vertices, we deduce that the dependence of the partition function on $p_{v_1}$ and $\gamma_{v_1}$ comes solely from the weight of third vertex in the first vertical line.
  
\begin{figure}[t!]
\centering
\includegraphics[width=155mm]{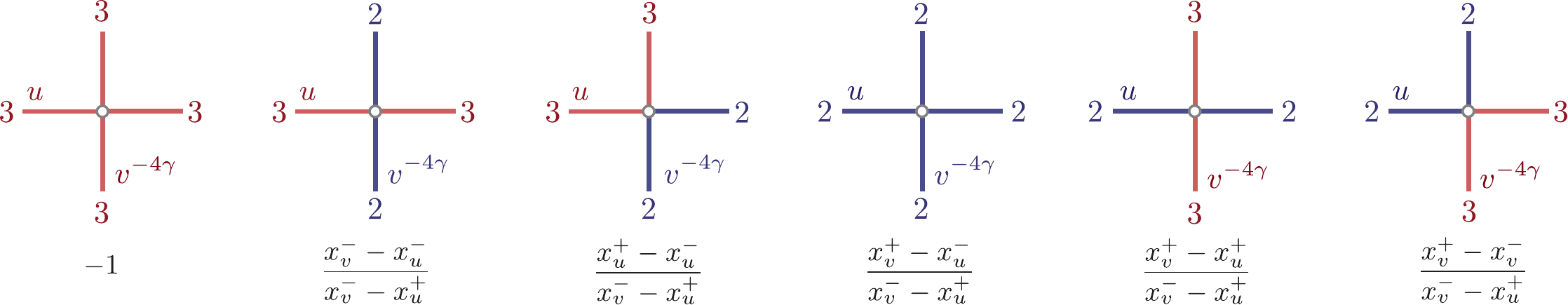}
\caption{The new six nonzero vertices. They are used to evaluate the  
domain wall partition function ${\mathcal{DW}}_N(\{ u \},\{ v \})$.  
\label{fig:sixvertexmodel2}}
\end{figure}

 Combining these observations with properties \ref{second} 
 and \ref{third}, we can determine the 
 dependence of the partition function 
 on $p_{u_k}, p_{v_k}$, $\gamma_{u_k}$ and $\gamma_{v_k}$  
 for every $k$ and it reads\footnote{The dependence 
 on the phases
$e^{i p_{u_i}}$ and $e^{i p_{v_i}}$ could be also derived 
using the map between the 
spin-chain frame and the string frame presented 
in Appendix \ref{Smatrix} since
the spin-chain frame $S$-matrix does not depend on 
those phases.}
 \begin{equation}
{\mathcal{P}}_N(\{ u \},\{ v \})
= 
\prod_{k=1}^N \left[-\frac{  \gamma_{v_k}}{\gamma_{u_k}} \,  e^{-\frac{i}{2} p_{v_k} \left(k-1/2\right)} e^{-\frac{i}{2} p_{u_k}\left(k-1/2-N\right)}  \right]\, 
{\mathcal{DW}}_N(\{ u \},\{ v \}) \, , 
\end{equation}
where ${\mathcal{DW}}_N(\{ u \},\{ v \})$ is a domain wall 
partition function with the same boundary conditions of    
figure \ref{fig:domainwall}, but with the six vertices of figure
\ref{fig:sixvertexmodel2} (it is possible this time to drop the $-4 \gamma$ from
 $v$ everywhere). 

Naturally, the domain wall partition function $\mathcal{DW}_N(\{ u \},\{ v \})$ inherits the properties of $\mathcal{P}_N(\{ u \},\{ v \})$ with small differences. We list them below,
\begin{enumerate}[I.]
 \item 
 ${\mathcal{DW}}_1(\{u_1\},\{v_1\}) = \frac{x^+_{u_1} - x^-_{u_1}}{x^-_{v_1} -
 x^+_{u_1}} \, ,$ \label{dfirst}
\vspace{2mm}
\item
${\mathcal{DW}}_N(\{ u \},\{ v \})
= - \frac{x^+_{v_{i+1}} - x^-_{v_i}}{x^-_{v_{i+1}}-
x^+_{v_i}}   \, 
{\mathcal{DW}}_N(\{ u  \},\{ v_1, \ldots, v_{i+1}, v_{i} \ldots , v_N \}) \, ,$ 
\vspace{2mm} \label{dsecond}
\item
${\mathcal{DW}}_N(\{ u \},\{ v \})
=- \frac{x^+_{u_{i+1}} - x^-_{u_i}}{x^-_{u_{i+1}}-
x^+_{u_i}}  \label{dthird}  \, 
{\mathcal{DW}}_N(\{ u_1 , \ldots , u_{i+1}, u_i, \ldots, u_N   \},\{ v \}) \, , $ 
\item 
${\mathcal{DW}}_N(\{ u \},\{ v \}) \Big|_{x^+_{u_N} = x^+_{v_N}}= \, \,    
(-1)^{N-1}\frac{x^+_{u_N} - x^-_{u_N}}{x^-_{v_N} - x^+_{u_N}} 
\, \, \prod_{i=1}^{N-1} \frac{x^+_{v_N} - x^-_{u_i}}{x^-_{v_N}- x^+_{u_i}} 
\, \, {\mathcal{DW}}_{N-1}(\{ u \},\{ v \}) \, .$ \label{dfour}
\end{enumerate}

Property \ref{dfirst} is again trivial and follows from 
the weight
of the third vertex of the figure \ref{fig:sixvertexmodel2}.
Properties \ref{dsecond} and \ref{dthird} are 
consequence of the Yang-Baxter equation and can 
be shown in a similar fashion 
using the procedure 
described in 
figure \ref{fig:domainwall}(b), but this time using the
 $R$-matrix built out of the vertices of 
 figure \ref{fig:sixvertexmodel2}.
Such $R$-matrix satisfies the unitary condition and the Yang-Baxter equation.

Property \ref{dfour} is a consequence of the
weights of the vertices. In the square lattice of figure \ref{fig:domainwall},
there are only two allowed vertices at the intersection of the lines 
$\{ u_N,v_N \}$. The vertex with weight proportional 
to $x^+_v - x^+_u$ is zero when $x^+_{u_N}=x^+_{v_N}$. So there is 
only one possible nonzero vertex at this intersection and it is 
not difficult to see that 
the lines corresponding to $u_N$ and $v_N$ are {\emph{frozen}} 
in this case and
the fourth property above follows. 

The solution for 
${\mathcal{DW}}_N(\{ u \},\{ v \})$ is given in a completely factorized form as follows\footnote{We were informed by O. Foda that the factorization of the S-matrix element has also been independently observed in an unpublished work by O. Foda and Z. Tsuboi.}
\begin{equation}
{\mathcal{DW}}_N(\{ u \},\{ v \}) = 
\prod^{N}_{i, j} \frac{1}{(x^-_{v_i} - x^+_{u_j})}
\prod_{i=1}^N (x^+_{u_i} - x^-_{u_i}) \prod_{j>i}^N (x^+_{v_j} - x^-_{v_i})  
\prod_{j>i}^N (x^+_{u_j} - x^-_{u_i}) \, .  
\label{eq:TheDomainFinalFormula}
\end{equation}

We will now prove that the solution given above is unique. The proof is by
induction. One can immediately see that the expression above satisfies
the property \ref{dfirst} above. In addition, inspecting the 
weights of the vertices of 
figure \ref{fig:sixvertexmodel2}, we see that the domain wall partition 
function has the form
\begin{equation}
{\mathcal{DW}}_N(\{ u \},\{ v \}) = (x^+_{u_N} - x^{-}_{u_N}) 
\prod_{i}^N \frac{1}{(x^-_{v_i} - x^+_{u_N})} \, g( \{u \} , \{ v \}) \, ,  
\end{equation}
where $g( \{u \} , \{ v \})$ is a polynomial of degree $(N-1)$ in $x^+_{u_N}$.
Suppose that ${\mathcal{DW}}_{N-1}$ is known. Using the property \ref{dsecond} and the 
result of property \ref{dfour}, we can derive recursion relations for $x^+_{u_N} =
x^+_{v_i}$ for $i = 1, \ldots , N$. These are $N$ conditions that uniquely fix $g( \{u \} , \{ v \})$
and consequently the domain wall partition function.  

The result for the domain wall partition function\footnote{Other instances where one can find factorized domain wall partition functions are \cite{Foda1} and \cite{Foda2}.}
 that we have 
just proven enables us to find an expression for the partition function
$\mathcal{P}_N(\{u\},\{v\})$ which is proportional to 
the matrix part of the hexagon 
form-factor of (\ref{eq:thehexagondyma}). Substituting $\mathcal{P}_N(\{u\},\{v\})$, 
we get that
\begin{equation}
h_{\Psi_1 \ldots \Psi_N | \bar\Psi_1 \ldots \bar\Psi_N} ( \{u \},\{v \}) 
= \frac{\prod_{i = 1}^{N} \prod_{j=1}^{N} 
h_{\Psi | \bar\Psi} ( u_i , v_j )}
{\prod_{i > j} h_{\Psi | \bar\Psi}(u_i, u_j) \prod_{i > j} 
h_{\Psi | \bar\Psi} (v_i, v_j)  }  \, .    
\label{eq:hexagonfatora}
\end{equation}

Using a similar reasoning, one can also derive the form-factor 
$h_{\bar\Psi_1 \ldots \bar\Psi_N | \Psi_1 \ldots \Psi_N} $. 
That  simply amounts to exchanging $\bar{\Psi} \leftrightarrow 
\Psi$ on the right hand side of the expression above.
This is the main result\footnote{The result  
(\ref{eq:hexagonfatora}) was derived in the string frame, however 
using the map between the spin-chain frame and the string frame it is
possible to show that it holds in the spin-chain frame as 
well. } of this section. In what follows, we 
proceed to the computation of the full three-point function. 

\subsubsection{The three-point functions} \label{full3pt2nonbps}
We consider now the full three-point 
function in the setup of figure \ref{tree3}, in which the 
excitations of the operator $\mathcal{O}_1$
and $\mathcal{O}_2$ are parametrized by the set of rapidities $\{ u \}$ and $\{ v \}$ 
respectively. We will be working in the asymptotic regime where 
all the lengths involved (both $L_i$ and $l_{ij}$) 
are large and all the finite size corrections can be neglected. 
According to the hexagon program, 
the asymptotic three-point function of these
operators at any loop order is given by
\begin{equation}
\left( \frac{C^{\rm{asym}}_{\bullet \bullet \circ}(N)}
{C_{ \circ \circ \circ }} \right)^2 =
\frac{\prod_{i=1}^N \, \mu_{\Psi}(u_i) 
\, \mu_{\Psi}(v_i) }{\prod\limits_{y=u,v} \Bigl({\rm{det}} \, \partial_{y_i} \phi_{y_j} \prod\limits_{i<j} S_{\mathfrak{su}(1|1)} (y_i, y_j)  \Bigr)} \times \mathcal{B}(N)^2 \, ,
\label{eq:ComNormalizacao}
\end{equation}
with
\begin{equation}
\mathcal{B}(N) =
\underset{\beta \cup \bar\beta=\{v\}}{
\underset{\alpha \cup \bar\alpha =\{u\}}{\sum}}
 (-1)^{X}\,w_{l_{13}}(\alpha , \bar\alpha) \,w_{l_{12}} (\beta , \bar\beta)
\, h_{\Psi_1 \ldots \Psi_{|\alpha|} | \bar{\Psi}_1 
\ldots \bar{\Psi}_{|\beta|}}(\alpha, \beta)\, h_{\bar\Psi_1 
\ldots \bar\Psi_{|\bar\beta|} | \Psi_1 \ldots \Psi_{|\bar\alpha|}}
(\bar\beta, \bar\alpha) \, .
\label{eq:2non1BPS}
\end{equation}
Moreover, $C_{ \circ \circ \circ }$ is the three-point
function of the three BPS operators obtained when $N=0$
and it is a constant combinatorial factor. The function
$\mu_{\Psi}(u)$ 
is the measure and as explained in 
\cite{hexagon} its 
square root gives the correct normalization of
the one-particle state in the hexagon program.
It is defined by 
\begin{equation}
\mu_{\Psi}(u) = \frac{i}{\underset{u=v}{{\rm{residue}}} 
\, \, h_{\Psi|\bar\Psi}(u,v)} \, . 
\label{eq:MeasureDefinition}  
\end{equation}
The phase $\phi_u$ in (\ref{eq:ComNormalizacao}) is
given by 
\begin{equation}
e^{i \phi_{u_j}} \equiv e^{i p_{u_j} L_1} 
\prod_{i \neq j}^{N_1} 
(- S_{\mathfrak{su}(1|1)}(u_j,u_k)) \, ,
\label{eq:defphi}
\end{equation}
and the phase $\phi_v$ is defined similarly. The determinant 
of the derivative of the phase $\phi_u$ is the usual  
Gaudin norm. 

The hexagon form-factors appearing in (\ref{eq:2non1BPS})
are evaluated in the 
spin-chain frame and they are nonzero only when 
$|\alpha|=|\beta|$. Moreover, 
$w_l$ are splitting factors, generically defined for a partition $\gamma \cup \bar{\gamma}$ of a set of rapidities $\{w\}$ by
\begin{eqnarray} \label{split}
w_{l}(\gamma, \bar\gamma) = \prod_{w_j \in \bar\gamma} 
\left( a_{l}(w_j) \prod_{w_i \in \gamma \, , \, i > j} 
S_{\mathfrak{su}(1|1)}(w_j, w_i) \right) \, ,
 \quad \quad {\rm with}  \quad \quad a_l (w) = e^{i p(w) l} \, .  
\end{eqnarray}
In the spin-chain frame normalization, the all-loop spin-chain
$\mathfrak{su}(1|1)$ $S$-matrix
is given by
\begin{equation}
S_{12}|_{{\rm{spin}}} \, | \chi_1^{3 \dot{1}} \chi_2^{3 \dot{1}} \rangle =
- (S^0_{12})^2 A_{12}|_{{\rm{spin}}} \, 
| \chi_2^{3 \dot{1}} \chi_1^{3 \dot{1}} \rangle
\equiv S_{\mathfrak{su}(1|1)}  | \chi_2^{3 \dot{1}} \chi_1^{3 \dot{1}} \rangle 
\, , \label{eq:DefinitionSsu11}
\end{equation}
where 
\begin{equation}
(S^0_{12})^2(u_1,u_2) = \frac{u_1 -u_2 + i}{u_1-u_2-i} 
\frac{(1-1/x_1^- x^+_2)^2}{(1-1/x_1^+x^-_2)^2} \frac{1}{
\sigma^2(u_1,u_2)} \, ,
\quad \quad A_{12}|_{{\rm{spin}}} = \frac{x_2^+ - x_1^-}{x_2^- - x_1^+} \, .
\nonumber
\end{equation}
The expression 
(\ref{eq:2non1BPS}) explicitly depends on the two lengths $l_{13}$ 
and $l_{12}$. It is possible to use 
the Bethe equations for the operator $\mathcal{O}_1$
(the unusual signs below appear because the excitations
are fermionic), 
\begin{equation}
a_{L_1}(u_j) \prod_{i \neq j}^{N_1} \left(- S_{\mathfrak{su}(1|1)}(u_j, u_i)\right) = 1 \, , 
\quad \quad j=1, \ldots, N_1 \, ,  
\label{eq:BetheAnyCoupling} 
\end{equation} 
and rewrite  it in terms
of the length $l_{12}$ only. After that, one gets at tree-level 
the scalar product of two off-shell 
$\mathfrak{su}(1|1)$ states.

In the expression (\ref{eq:2non1BPS}) above, $(-1)^X$ accounts for some sign differences between the two hexagons involved in the structure constant. It was already noticed in \cite{hexagon}, that such signs were important in order to get a match with both weak and strong coupling data. The empirical rule found there was to include the factor  $(-1)^M$, where  $M$ is nothing but the total number of magnons of the second hexagon (equivalently $M=|\bar\alpha| + |\bar\beta| $). 
In a similar way, we have found the need of introducing 
additional signs to get an agreement with the tree-level data. In total, 
we have that 
\begin{equation}
X=
|\bar{\alpha}| N \, .   
\end{equation}
Note that $M$  should be always  even in order to get a nonzero hexagon so that $(-1)^M$ does not introduce any sign.

The two particle fermionic hexagon form-factor is related to $S_{\alg{su}(1|1)}$.  By explicitly evaluating the hexagon form 
factors for $N=1$ in the spin-chain frame, one can check 
that the following identity holds
\begin{equation}
\frac{h_{\Psi | \bar\Psi}(u,v)}{h_{\Psi | \bar\Psi}(v,u)}
= 
S_{\mathfrak{su}(1|1)}(u,v) .
\end{equation}
This identity reflects the Watson equation for form-factors which is, by construction, automatically satisfied by the hexagon ansatz. Using 
this relation, we can write the three-point function in a more concise way. Given two sets $\rho_u = \{ u_1, \dots, u_{|\rho_u|}\}$ and $\rho_v = \{ v_1, \dots, v_{|\rho_v|}\}$, let us introduce the notation
\begin{gather}
e^{i p_{\rho_u} l} = \prod_i e^{i p(u_i)\, l}\,,\,\,\,\,\,
h_{\Psi|\bar{\Psi}} (\rho_u ,\rho_v) = \prod_{i,j} h_{\Psi |\bar{\Psi}} (u_i ,v_j) \,,\\
 h^{>}_{\Psi|\bar{\Psi}} (\rho_u ,\rho_v) = \prod_{i>j} 
 h_{\Psi |\bar{\Psi}} (u_i ,v_j) \,,\,\,\,\,\, 
 h^{<}_{\Psi|\bar{\Psi}} (\rho_u ,\rho_v) = 
 \prod_{i<j} h_{\Psi |\bar{\Psi}} (u_i ,v_j)\,.
\end{gather}
The equation (\ref{eq:2non1BPS}) can then be rewritten as
\begin{equation}
\begin{aligned} \label{concise2nonbps}
\mathcal{B}(N)= \hspace{-1mm}
 \underset{\beta \cup \bar\beta=\{v\}}{
\underset{\alpha \cup \bar\alpha =\{u\}}{\sum}}(-1)^X\,
\frac{  e^{ip_{\bar\alpha} l_{13}+ip_{\bar\beta} l_{12}} \,\,
h^{<}_{\Psi |\bar\Psi}(\bar\alpha,\alpha) 
h^{<}_{\Psi |\bar\Psi}(\bar\beta,\beta) 
h_{\Psi|\bar\Psi}(\alpha,\beta) 
h_{\bar\Psi|\Psi}(\bar\beta,\bar\alpha)    }
{h^{>}_{\Psi |\bar\Psi}(\alpha,\bar\alpha) 
h^{>}_{\Psi |\bar\Psi}(\beta,\bar\beta) 
h^{>}_{\Psi|\bar\Psi}(\alpha,\alpha) 
h^{>}_{\Psi|\bar\Psi}(\beta,\beta)  
h^{>}_{\bar\Psi|\Psi}(\bar\alpha,\bar\alpha) 
h^{>}_{\bar\Psi|\Psi}(\bar\beta,\bar\beta)  }\,.
\end{aligned}
\end{equation}
Let us now further expand on the comparison with data. In subsection 
\ref{sec:Thedeterminant}, a determinant expression for the three-point function
of three generic $\mathfrak{su}(1|1)$ operators at 
tree-level was derived, see (\ref{determinantI}) and (\ref{determinant}).
 This result, more precisely $C_{123}/ 
\mathcal{R}$ with $N_3=0$ and with a suitable normalization of
the wave-functions,   
can be compared with the
tree-level limit of 
$\mathcal{B}(N)$.
One way of finding the relevant normalization of the 
wave-functions
is by comparing the two results for the simplest case 
$N=1$.  
In this section, all the hexagon form-factors are evaluated in the
spin-chain frame and at order $g^0$, one has 
\begin{equation}
h_{\Psi | \bar\Psi} (u, v) = 
-h_{ \bar\Psi | \Psi }(u, v) = 
\frac{1}{u-v} \; , \quad \quad 
S_{su(1|1)}(u_i, u_j) = -1 \, , 
\end{equation}
and 
\begin{equation}
\mathcal{B}(1)= \frac{1}{u-v} \, (1 - 
e^{i p(v) l_{12}} \, e^{i p(u) l_{13}}) \, .
\end{equation}
Using the Bethe equation for the operator $\mathcal{O}_1$, it 
is not
difficult to see that the result above 
agrees with the result for
$C_{123}/ 
\mathcal{R}$ given in (\ref{determinantI}) 
if we multiply 
this later by the normalization factor $\mathcal{N}(u)\times \mathcal{N}(v)$ where $\mathcal{N}$ is given by
\begin{equation}
\mathcal{N}(u)=
\sqrt{i}\,(e^{- i p(u)} -1)  \, .
\end{equation}
In this way, we have found the correct normalization 
of the wave-functions
to compare the two results. We should then multiply the 
wave-function 
given in (\ref{wavefunction}) by these normalization factors for all
rapidities. One can now evaluate 
$\mathcal{B}(N)$ for different
values of $N$ and check that in fact it reproduces the results
obtained from the determinant formula. Alternatively, 
one can directly compare the complete $C_{123}$ given in 
(\ref{determinantI}) computed with 
standard normalized wave-functions with the properly 
normalized structure constants computed with the hexagon
program 
as in (\ref{eq:ComNormalizacao}).

We have seen that the 
factor $\mathcal{B}(N)$ of the tree-level structure constant 
can be written as a determinant, which is directly related to the fact that the scalar product
of two off-shell $\mathfrak{su}(1|1)$ states can also be 
written 
in the form of a 
determinant, see also \cite{offsu,offsu2}. 
This property appears to be special to $\mathfrak{su}(1|1)$ and it is currently not known if such determinant expressions exist in the other
rank one sectors, namely $\mathfrak{su}(2)$
and $\mathfrak{sl}(2)$.
A natural question is whether 
$\mathcal{B}(N)$
can still be written as a determinant (or in another computationally efficient form) when loop corrections are included. 
We have not found a full answer to this question, but in what follows we will show that at strong coupling leading order such simplification exists and the result can be indeed expressed in the form of a determinant.

\paragraph{Strong coupling limit} As a prediction for 
a direct strong coupling computation of 
the asymptotic three-point functions 
considered 
in this section, we consider the large coupling 
limit of (\ref{concise2nonbps}). There are several regimes 
in the kinematical space and here we 
focus on the so-called BMN regime for which the 
momentum  scales as $p \sim 1/g$ and the rapidities
scales as $u \sim g$. 
Using that in this regime
\begin{equation}
x_u^{\pm} = x_u \pm \frac{i\, x_u\,p_u}{2}  +\mathcal{O}(1/g^2)\,,
\end{equation}
and the leading expression for the dressing phase, i.e.
the AFS dressing factor of \cite{AFSdressing,AFSdressing2}, it 
is simple to derive that
\begin{equation}
h_{\Psi | \bar\Psi} (u, v) =   
\frac{\sqrt{p_u \,p_v \,x_u x_v}}{x_u - x_v} + 
\mathcal{O}(1/g^2)\,.
\label{eq:strongh}
\end{equation}
When we plug this expression in (\ref{eq:2non1BPS}) and 
use the fact that 
\begin{equation}
S_{\mathfrak{su}(1|1)}(u,v) \simeq -1+\mathcal{O}(1/g) \, ,
\quad \quad {\rm{and}} \quad \quad L_i \sim g \, , 
\label{eq:strongS}
\end{equation}
where the condition on $L_i$ is necessary  
in order for the operators to satisfy the Bethe equations  
(\ref{eq:BetheAnyCoupling}), we obtain after a little massaging that the factor 
$\mathcal{B}(N)$ contributing to 
the strong coupling structure constant 
can be expressed as
 \begin{equation}
\begin{aligned}
\mathcal{B}(N)= & \sqrt{ \prod_{i=1}^{N} p_{u_i}  p_{v_i} x_{u_i} x_{v_i} }  
 \underset{\beta \cup \bar\beta=\{v\}}{
\underset{\alpha \cup \bar\alpha =\{u\}}{\sum}}  (-1)^{X}\, 
\, (-1)^{P_{\alpha}+P_{\beta}} \,e^{i p_{\bar\alpha} l_{13} +i p_{\bar\beta} l_{12}}   \, \frac{ g^{\alpha \alpha}_{>}\, g^{\beta \beta}_{>}\,g^{\bar\alpha \bar\alpha}_{>}\, g^{\bar\beta\bar\beta}_{>}}{g^{\alpha \beta} g^{\bar\alpha \bar\beta}} \,,
\end{aligned}
 \end{equation}
 where $(-1)^{P_{\alpha}}$ is defined as the sign of permutation of the ordered set $\{u\}$ which gives $\alpha \cup \bar\alpha $. In this expression, we use that $g^{\alpha \beta } = \underset{u_i\in\alpha, v_j \in \beta}{\prod}(x_{u_i} -x_{v_j})$ and $g^{\alpha \alpha }_{>} = \underset{i>j}{\underset{u_i,u_j\in\alpha }{\prod}}(x_{u_i} -x_{u_j})$\,.
 
 This formula can be finally recasted as the following determinant 
 \begin{equation}
\mathcal{B}(N)= 
  (-1)^{\frac{N (N-1)}{2}}   
  \sqrt{ \prod_{i=1}^{N} p_{u_i}  p_{v_i} x_{u_i} x_{v_i} } \,
  \det_{1\leq i,j \leq N} \left[ 
  \frac{ 1-e^{i p_{u_i} l_{13} + i p_{v_j}l_{12}}}{x_{u_i} - x_{v_j}}
  \right]\,.
 \end{equation} 
To compute the properly 
normalized structure constant of (\ref{eq:ComNormalizacao}) 
in the strong coupling limit, we also need to find the 
leading contribution both of the measure $\mu_{\Psi}(u)$ and 
the
Gaudin norm at large coupling. 
Using the result (\ref{eq:strongh}) for 
$h_{\Psi|\bar\Psi}(u,v)$ 
and the definition of the measure 
$\mu_{\Psi}(u)$ of (\ref{eq:MeasureDefinition}), it is
not difficult to see that
\begin{equation}
\mu_{\Psi}(u) =  i \frac{\partial_u x_u}{p_u x_u} + \mathcal{O}(1/g) \, .
\end{equation}

The Gaudin norms can be evaluated using the definition
of the phases $\phi_{u_j}$ given in (\ref{eq:defphi}) and
the $S_{\mathfrak{su}(1|1)}(u,v)$ of (\ref{eq:strongS}), 
leading to
\begin{equation}
{\rm{det}} \, \partial_{u_i} \phi_{u_j} = \prod_{i=1}^N 
\partial_{u_i} p_{u_i} L_1 + \mathcal{O}(1/g^{N+1}) \, ,
\end{equation}
and the result for  
${\rm{det}} \, \partial_{v_i} \phi_{v_j}$ 
is analogous
to the one above
with both $u_i$ and $L_1$ replaced by $v_i$ 
and $L_2$ respectively. 

The strong coupling limit of the 
structure constants $C^{\rm{asym}}_{\bullet \bullet \circ}(N)$
of (\ref{eq:ComNormalizacao}) can then be 
obtained by assembling together 
these
results. 
By analyzing how these several contributions\
scale 
with $g$,
it follows that the structure constants are of order $\mathcal{O}(1)$ in the coupling
for any $N$.

\subsection{The 3 non-BPS case}\label{3nonBPSsection}
\begin{figure}[t!]
\centering
\includegraphics[width=140mm]{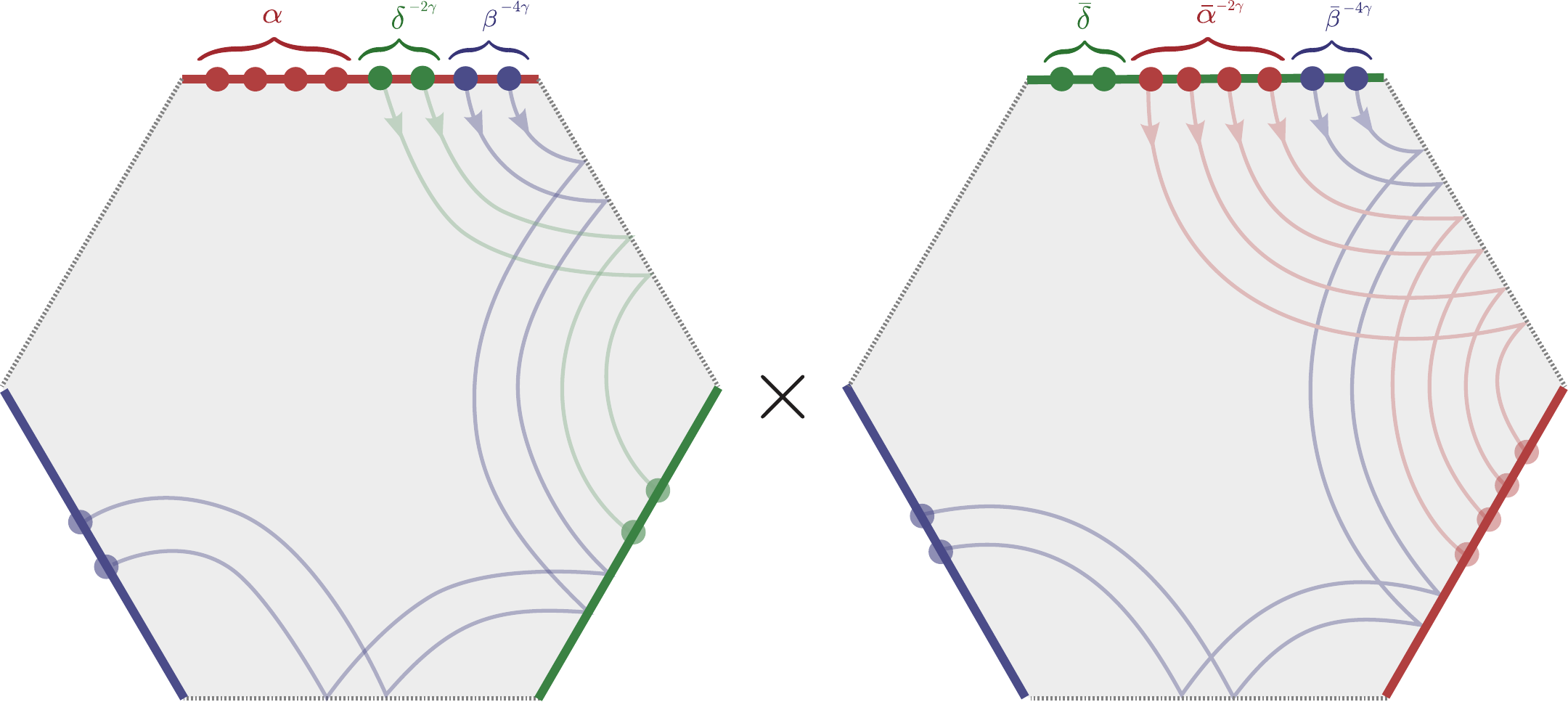}
\caption{The product of two hexagons is the core of 
the three-point function. We divide each set of 
excitations in three partitions and distribute them over 
the two form-factors. Two out of the three sets are 
mirror transformed which is equivalent to transfer the 
excitations to the other physical edges.
\label{3nonbps}}
\end{figure}
In this subsection, we will compare the results for 
the three-point functions of three $\mathfrak{su}(1|1)$ non-BPS
operators obtained in \cite{CaetanoFleury} at one-loop order
by a direct perturbative calculation 
with the results predicted by the hexagon program. 
This constitutes a rather nontrivial test of the hexagon 
program. 

In \cite{CaetanoFleury},  we have considered a setup  
consisting of three operators in the $\mathfrak{su}(1|1)$, 
where $\mathcal{O}_1$ 
was made out of 
 $Z \equiv \Phi^{34}$ and 
$\Psi \equiv \psi_{\alpha=1}^{4}$ 
and $\mathcal{O}_2$ was made out of the corresponding 
conjugate fields $\bar{Z} = (Z)^{*}$ and 
$\bar{\Psi} \equiv (\Psi)^{\dagger}$. The third 
operator $\mathcal{O}_3$ was chosen to be a 
certain rotated operator in order to have a non-extremal 
three-point function. More specifically
\begin{equation}
\mathcal{O}_3=\frac{1}{ (L_3-N_3)!^2}\sum_{1 \le n_1 < \ldots<n_{N_3}
\le L_3} \psi^{(3)}_{n_1,n_2,\ldots,n_{N_3}} \, (\alg{R}^{2}_{\;\;4 } 
\alg{R}^{1}_{\;\;3 })^{L_3-N_3}  \cdot 
 \Tr \, ( Z \dots \underset{n_1}{\Psi} \dots 
\underset{n_2}{\Psi} \dots Z )\ \,, 
\end{equation}
where  $\psi^{(3)}$
is the wave-function depending on the momenta of the 
excitations $\Psi$. Here  $\alg{R}^{a}_{\; \; b}$ are the $\mathfrak{su}(4)$ generators with 
$a,b=1, \ldots, 4$. For all operators $\mathcal{O}_i$, $L_i$ and $N_i$ are the corresponding length and number of excitations.

At one-loop level, the corrections coming from both the 
wave-functions and Feynman diagrams were 
computed in \cite{CaetanoFleury}. This latter correction turned 
out to be encoded in the form of some splitting operators 
to be inserted on top of the tree-level contractions. When combined both corrections together we have found a remarkably simple factorized result given by

\begin{equation}
C_{\bullet \bullet \bullet}
=\mathcal{C}\;\frac{\prod\limits_{k=1}^{3}\prod\limits_{\substack{i< j}}^{N_k}f(y^{(k)}_i,y^{(k)}_j)}{\prod\limits_{i=1}^{N_1}\prod\limits_{j=1}^{N_2}f(y^{(1)}_i,y^{(2)}_j)}\prod\limits _{k=1}^{N_1} \left[1-(y^{(1)}_{k})^{L_2}\prod\limits_{i=1}^{N_2} \left(-S(y^{(2)}_i,y^{(1)}_k)\right)\right]\,, \label{oneloop}
\end{equation}
where we are using the notation $y^{(i)}_k=e^{i p^{(i)}_k}$, with $\{p^{(i)}_k\}_{k=1}^{N_i}$  being the set of momenta characterizing the excitations of the operator $\mathcal{O}_i$. The normalization factor $\mathcal{C}$ and $f$ are given by
\begin{equation}
\begin{aligned}
\mathcal{C}&=\prod_{i=1}^3 \left(\frac{L_i}{\mathcal{N}^{(i)}}\right)^{1/2}\left[1+g^2 \left(N_3^2-1\right)-\frac{1}{4}\sum_{i=1}^{3}\bold\gamma_i \right]\,, \label{normalfactor} \\[10pt]
f(s,t)&=(s-t)\left[1-\frac{g^2}{2}\left(\frac{s}{t}+\frac{t}{s}-\frac{1}{s}-s-\frac{1}{t}- t +2\right)\right]\,.
\end{aligned}
\end{equation}
with $\gamma_i$ being the anomalous dimension of the operator $\mathcal{O}_i$.

In order to compare the perturbative calculations 
with the results of
the hexagon program, 
we have to properly 
normalize the wave-functions
$\psi^{(i)}$. One way of finding the correct normalization is 
to use the results of the previous subsection for 
two non-BPS operators 
when
$N=1$ and match it with the corresponding one-loop three-point 
function. Since the wave-functions only contain 
local information 
of each operator, they ought to be the same 
for any three-point 
function within the same sector.
In order to compute the three-point function of 
one BPS and two 
non-BPS operators at one-loop we make use
of the splitting insertions for fermions obtained in 
\cite{CaetanoFleury}. Once the comparison with
$\mathcal{B}(1)$ of
(\ref{eq:2non1BPS}) at one-loop order is made,
one finds that the two results agree if the one excitation 
wave-function
is normalized as
\begin{equation}
\psi(n_1) = \mathcal{N}(p) \, e^{i p n_1} \, , \quad {\rm{with}}
\quad \quad \mathcal{N}(p)= 
\sqrt{i} \,\frac{(e^{- i p} -1)}{1+ g^2 (e^{ip}+e^{-ip}-2)} 
\, . \label{eq:OneLoopNormalization}
\end{equation}
In the case of more than one 
excitation the normalized wave-function is obtained 
by multiplying it
by $\mathcal{N}(p_i)$ given above for 
all the excitations $i$.

We want now to access this three-point function 
within the framework of the hexagon program. In order to match our previous setup, we choose the set of excitations as follows: 
 the physical edge associated to the operator $\mathcal{O}_1$ contains $N_1$ excitations of type
$\Psi = \chi^{ 3 \dot{1}}$, the edge  corresponding to $\mathcal{O}_2$ has 
$N_2$ excitations of the type  
$\bar\Psi=\chi^{2 \dot{4}}$ 
and remaining physical edge has $N_3=l_{23}=N_2-N_1$ excitations of type
$\Psi = \chi^{ 3 \dot{1}}$. For details about the 
construction of operators in the hexagon formalism, we refer 
the reader to the Appendix B of \cite{hexagon}. The relevant 
hexagon form-factor to be considered contains three sets of type of excitations. Accounting for the mirror transformations illustrated in figure \ref{3nonbps}, and given three generic sets of rapidities $\{u_i\}_{i=1}^{N_u},\{w_i\}_{i=1}^{N_w}$ and $\{v_i\}_{i=1}^{N_v}$ it reads
\begin{equation}
\mathfrak{h}_{\{u\},\{w\},\{v\}}\equiv (-1)^{N_v+N_w} h{\overbrace{^{3\dot{1},\ldots,3\dot{1},}}^{N_u}\overbrace{^{1\dot{3},\ldots,1\dot{3},}}^{N_w}\overbrace{^{2\dot{4},\ldots,2\dot{4}}}^{N_v}}(u_1,\ldots,w^{-2\gamma}_1,\ldots,v^{-4\gamma}_1,\ldots),
\end{equation}
where we are using the notations of (\ref{eq:hexagonform}) and the sign $(-1)^{N_v+N_w}$ comes from the crossing rules for the excitations as described in Appendix \ref{Smatrix}.
The full asymptotic three-point function is then built 
out of this hexagon form-factor  through

\begin{equation}
\left( 
\frac{C^{\rm{asym}}_{\bullet \bullet \bullet}(N_1,N_3)}
{C_{\circ \circ \circ}} \right)^2
=
\frac{\prod\limits_{i=1}^{N_1} \, \mu_{\Psi}(u_i) 
\, \prod\limits_{i=1}^{N_2} \, \mu_{\Psi}(v_i) 
\, \prod\limits_{i=1}^{N_3} \, \mu_{\Psi}(w_i) }{
\prod\limits_{y=u,v,w} \Bigl({\rm{det}} \, \partial_{y_i} \phi_{y_j} \prod\limits_{i<j} S_{\mathfrak{su}(1|1)} (y_i, y_j)  \Bigr)
} \,\times\mathcal{C}(N_1,N_3)^2 \, ,
\label{eq:ProperlyNormalized3}
\end{equation}
where $C_{\circ \circ \circ}$ is a constant 
combinatorial factor equal to the three-point function
of the three BPS operators obtained when $N_1=N_3=0$.
The functions $\mu_{\Psi}(u)$ and 
$\phi_{u_i}$ were defined in (\ref{eq:MeasureDefinition})
and (\ref{eq:defphi}) respectively. 
Finally,

\begin{equation}
\mathcal{C}(N_1,N_3)= \underset{\delta \cup \bar\delta =\{w\}}{\underset{\beta \cup \bar\beta=\{v\}}
{\underset{\alpha \cup \bar\alpha =\{u\}}{\sum}}}
 (-1)^{X}\,w_{l_{13}}(\alpha , \bar\alpha) \,w_{l_{12}} (\beta , \bar\beta)\, w_{l_{23}} (\delta , \bar\delta)
\, \mathfrak{h}_{\alpha,\delta,\beta}\, 
\mathfrak{h}_{\bar{\delta},\bar\alpha,\bar\beta} \, , 
\label{eq:3non}
\end{equation}
with the splitting factors given in (\ref{split}).
Upon expanding 
 $\mathcal{C}(N_1,N_3)$ above up to one-loop we have 
found 
that it matches with the properly 
normalized\footnote{Equivalently, one can compare
the data given in (\ref{oneloop})  using a
standard normalized wave-functions, i.e. $\psi(n_1)$ in
(\ref{eq:OneLoopNormalization}) with $\mathcal{N}(p)=1$, 
with the three-point function 
obtained using 
the hexagon program given in 
(\ref{eq:ProperlyNormalized3}) including the prefactor in front of $\mathcal{C}(N_1,N_3)^2$.}  results referred to above, 
once $X$ is taken to be\footnote{We point out that this choice for $X$ is not unique with the amount of data we fitted. A more thorough study with a larger number of excitations might narrow the space of solutions for $X$.}
\begin{equation}
X =  {|\bar\delta| N_1+ |\bar\alpha| N_2 +|\beta| N_3}\,.
\end{equation}
Note that once again this differs from the rule advocated in \cite{hexagon} and mentioned in the previous section. It is  desirable to have a deeper understanding of the origin of these relative signs between the hexagons.

\section{Conclusions}

In this paper, we have studied the three-point functions of 
operators 
in the $\mathfrak{su}(1|1)$ sector, i.e., containing a single type of fermionic excitations. We 
have managed to parametrize the most general 
configuration 
in this rank one sector 
by a sort of polarization vectors and showed that 
at leading order the structure constant\footnote{There is a single conformally invariant 
tensor structure for any of 
these configurations \cite{CaetanoFleury}.} can be expressed 
in the form of a determinant. In a particular limit, such 
determinant reduces to an off-shell 
scalar product of $\mathfrak{su}(1|1)$ Bethe states.

We have then applied the hexagon program of \cite{hexagon} to 
study 
all-loop correlators in this sector. We have started with 
the case of 
one BPS 
and two non-BPS operators. We have 
shown that the relevant hexagon form-factor 
can be identified with a domain wall partition function 
of a six-vertex model defined by some entries of 
the $\mathfrak{su}(2|2)$ 
$S$-matrix. This 
property appears 
to be specific for this sector and in particular, it is 
no longer 
true for other rank one sectors,
where only at tree-level 
such identification can be made. 
A peculiar 
feature of the domain wall partition function we have found 
here is that it completely 
factorizes, see (\ref{eq:hexagonfatora}), and its 
computation becomes rather economical. We then assembled 
a pair of such completely factorized hexagon form-factors 
to compute the structure constants. Upon expanding it at 
leading order in the coupling constant we have checked 
that it matched precisely with our tree-level prediction 
once we include a relative sign factor between the 
two hexagons. This is an addition to the prescription 
put forward in \cite{hexagon}, where it was already 
noticed the need of including some relative signs 
when the two hexagons are multiplied. This particular 
point certainly needs a clarification. 
The expression for the structure constants is given in
(\ref{concise2nonbps}) and one interesting limit of this
expression is the strong coupling limit which
is a prediction for a future string theory 
computation. We showed that in the BMN regime the
structure constant admits surprisingly a determinant
expression for any number of excitations.
An interesting 
future direction that comes out of our 
results is to investigate the possibility of 
writing the full three-point function at finite coupling in a way that 
circumvents the computationally costly sums over partitions of Bethe roots. This is generally hard but within this particular setup where the hexagon form-factors are explicitly known, it might be a good starting point. Equally interesting is to
take the classical limit of
our result, for $L_1,L_2,N_1,N_2 \rightarrow \infty $ with
$L_i/N_i$ fixed. Such limit for operators within 
the $\mathfrak{su}(2)$ 
sector was recently considered in \cite{ShotaLast}.

We finally studied a particular configuration of three non-BPS operators in the same setup previously studied in \cite{CaetanoFleury} up to one-loop. We have managed to check that the structure constant computed from the hexagon program nicely reproduces the perturbative data of \cite{CaetanoFleury} once  we include some relative signs between the two hexagons. This additional feature is analogous to the previous case. The one-loop structure constants computed in \cite{CaetanoFleury} have a completely factorized form even at one-loop. This raises hopes that it might be possible to find an all-loop simplification coming out of the hexagons. We hope to address this question in the future.

\section*{Acknowledgements}

We would like to thank S. Komatsu and P. Vieira  
for many central discussions and suggestions.  We
also thank B. Basso, O. Foda, V. Gon\c{c}alves and 
H. Nastase for 
discussions and for carefully reading the draft and 
also I. Kostov and D. Serban
for discussions.
The work of 
JC was supported by European Research 
Council (Programme ``Ideas'' ERC- 2012-AdG 320769 AdS-CFT-solvable).
TF would like to thank the warm hospitality of
the Perimeter Institute where this work was initiated. TF 
would like to
thank FAPESP
grants 13/12416-7 and 15/01135-2 for financial support.

\appendix
\section{The String Frame $\mathfrak{su}(2|2)$-invariant $S$-matrix}  \label{Smatrix}

In this Appendix, we set our conventions for the string frame 
$\mathfrak{su}(2|2)$-invariant $S$-matrix.
As explained in the main part of the paper,
we evaluate the hexagon form-factors in the string frame and 
use the map between the frames 
to translate the results to the spin-chain frame when comparing with
the available data. The string frame $S$-matrix obeys the standard
Yang-Baxter equation and its action on the states does not   
produce $Z$ markers. The $S$-matrix 
has the following nonzero matrix elements 
($\epsilon^{12}=\epsilon_{12}=1$) 
\begin{align} 
& S_{12} | \chi^a_1 \chi^b_2 \rangle = A_{12} 
| \chi^{ \{a }_2 \chi^{ b \}}_1 \rangle +  B_{12} 
| \chi^{ [ a }_2 \chi^{ b ]}_1 \rangle + \frac{1}{2} C_{12} 
\epsilon^{ab} \epsilon_{\alpha \beta} | \chi_2^{\alpha} \chi_1^{\beta} 
\rangle \, , \nonumber & \\[0.6em] 
& S_{12} | \chi^{\alpha}_1 \chi^{\beta}_2 \rangle = D_{12} 
| \chi^{ \{\alpha }_2 \chi^{ \beta \}}_1 \rangle +  E_{12} 
| \chi^{ [ \alpha }_2 \chi^{ \beta ]}_1 \rangle + \frac{1}{2} F_{12} 
\epsilon^{\alpha \beta} \epsilon_{ a b } | \chi_2^{ a } \chi_1^{ b } 
\rangle \, , \nonumber & \\[1em]
& S_{12} | \chi^{ a }_1 \chi^{\alpha}_2 \rangle = G_{12} 
| \chi^{ \alpha }_2 \chi^{ a }_1 \rangle +  H_{12} 
| \chi^{ a  }_2 \chi^{ \alpha }_1 \rangle \, , \hspace{20mm} \nonumber &
\\[1em]
& S_{12} | \chi^{ \alpha }_1 \chi^{ b }_2 \rangle = K_{12} 
| \chi^{ \alpha }_2 \chi^{ b }_1 \rangle +  L_{12} 
| \chi^{ b  }_2 \chi^{ \alpha }_1 \rangle \, , \hspace{21mm}  & \label{saction}
\end{align}
where using the definitions 	
\begin{equation}\label{etass}
\eta_i=\eta(x_i^+, x_i^-) = e^\frac{i p_i}{4} \sqrt{i(x^{-}_i - x^{+}_i)}
\, , \quad \quad \tilde\eta_1= \eta_1 \, 
e^\frac{i p_{2}}{2} \, , \quad \quad \tilde\eta_2= \eta_2 \, 
e^\frac{i p_1}{2} \, ,    
\end{equation}
the matrix elements are 
\vspace{2mm}
\begin{align}
  & A_{12} = \frac{x_2^+ - x_1^-}{x_2^- - x_1^+} 
\frac{\tilde\eta_2 \eta_1}{\eta_2 \tilde\eta_1 } \, ,  
& \hspace{10mm} &\\[1em]
  & B_{12}=\frac{x_2^+ - x_1^-}{x_2^- - x_1^+} \left(1 - 2 
\frac{1 - 1/x_2^- x_1^+}{1-1/x_2^+x_1^+} 
\frac{x_2^- - x_1^-}{x_2^{+} - x_1^-}\right)
\frac{\tilde\eta_2 \eta_1}{\eta_2 \tilde\eta_1 } \, , 
& \hspace{10mm} & \\[1em]
 & C_{12} = - \frac{2 \eta_1 \tilde\eta_2}{i x_1^+ x_2^+} 
 \frac{1}{1 - 1/x_1^+ x_2^+} \frac{x_2^- - x_1^-}{x_2^- - x_1^+} \, , & 
 \hspace{10mm} & \\[1em]
  & D_{12} = -1 \, , 
& \hspace{10mm} & \\[1em]
& E_{12} = - \left( 1 - 2 \frac{1 -1/x^+_2 x_1^-}{1 - 1/ x^-_2 x^-_1}
\frac{x_2^+ - x_1^+}{x_2^- - x^+_1} \right) \, , 
& \hspace{10mm} & \\[1em] 
& F_{12} = \frac{2 i (x_1^+ - x_1^-)(x_2^+ - x_2^-)}
{\tilde\eta_1 \eta_2 \, x_1^- x_2^-} \frac{1}{1 - 1/ x_1^- x_2^-} 
\frac{x_2^+ - x_1^+}{x_2^- - x_1^+}  \, , 
& \hspace{10mm} & \\[1em]
&  G_{12} = \frac{x_2^+ - x_1^+}{x_2^- - x_1^+} 
\frac{\eta_1}{\tilde\eta_1} \, , 
& \hspace{10mm} & \\[1em]
&  H_{12} = \frac{x_2^+ - x_2^-}{x_2^- - x_1^+} \frac{\eta_1}{\eta_2}  
\, , 
& \hspace{10mm} & \\[1em]
&  K_{12} = \frac{x_1^+ - x_1^-}{x_2^- - x_1^+} 
 \frac{ \tilde\eta_2}{\tilde\eta_1}   
\, , 
& \hspace{10mm} & \\[1em]
& L_{12} = \frac{x_2^- - x_1^-}{x_2^- - x_1^+}  
\frac{\tilde\eta_2}{\eta_2} 
\, .
& \hspace{10mm} & 
\end{align}

One of the reasons
to evaluate the hexagon form-factors
using the string frame $S$-matrix is the fact that
all the branch 
cut ambiguities of the function $\eta_i$ 
can be resolved by
the variable $z$ parametrizing the rapidity torus \cite{RapidityTorusI,Foundations,RapidityTorusII}.
This means also that there is no ambiguities when one performs
crossing transformations. 
The way the variables transform will
be explained in the next subsection. 
Using these crossing transformations, we have checked that
the hexagon form-factors have all the expected properties 
such as 
invariance under cyclic rotations and 
consistency between all possible ways to move all the 
particles to a single physical edge. 
Using the variable $z$ corresponds 
to choosing a branch for the
square roots in $\eta_i$.

\subsection{Mirror transformations of fermions}

The prescription to evaluate the hexagon form-factor 
in the case where 
not all the physical excitations are in one edge 
is to perform crossing
transformations and move all of them to a single physical edge.
The string frame $\mathfrak{su}(2|2)$ $S$-matrix is a meromorphic 
function when written in terms of a complex coordinate $z$ parametrizing
the rapitidy torus.
A crossing transformation, denoted by $2 \gamma$ in what follows,
corresponds to shift $z$ by half of the imaginary
period of the torus. The transformation of the matrix elements of
the $S$-matrix under crossing can be deduced using the following
transformations 
\begin{equation} \label{crossx}
\eta_i^{- 2 \gamma} = - \frac{i \eta_i}{x_i^{+}} \, , \quad \quad 
\eta_i^{2 \gamma} =  \frac{i \eta_i}{x_i^{+}} \, , \quad  \quad
(x_i^{-})^{\pm 2 \gamma}  
= \frac{1}{x_i^{-}} \, , \quad \quad (x_i^{+})^{\pm 2 \gamma} 
= \frac{1}{x_i^{+}} \, ,  
\end{equation}
where $\eta_i$ is defined in (\ref{etass}). In addition, 
it is also necessary to
know how the fundamental excitations transform under crossing.
According
to Appendix D of \cite{hexagon}, the fundamental excitations 
decompose as follows under the diagonal $\mathfrak{psu}(2|2)_D$ symmetry 
preserved by the hexagon
\begin{equation}
\chi^{A \dot{B}}(u) \, \,  \mapsto \, \,  
\chi^{A}_D(u) \, \chi^{\dot{B}}_D(u^{-2 \gamma}) \, . 
\end{equation}
Moreover, one can also find in the Appendix D of \cite{hexagon} the 
relations
\begin{equation}
\chi^{a}(u^{2 \gamma}) = - \chi^{a} (u^{-2 \gamma}) \, , \quad \quad 
\chi^{\alpha}(u^{2 \gamma}) = \chi^{\alpha}(u^{-2 \gamma}) \, .
\end{equation}
Using the above equations, one can deduce the transformations of
the fundamental excitations. For example, one has\footnote{We thank Shota Komatsu for 
informing us about the transformations 
of the fermionic excitations}
\begin{eqnarray} \label{crossrules}
\chi^{\alpha \dot{a}} \, \overset{2 \gamma}{\rightarrow} 
\, \chi^{a \dot\alpha} \, , \, \, 
\quad \chi^{a \dot\alpha}  \, \overset{2 \gamma}{\rightarrow} \,
- \chi^{\alpha \dot{a}} \, , \, \, \quad
\chi^{\alpha \dot{a}} \, \overset{-2 \gamma}{\rightarrow} 
\, - \chi^{a \dot\alpha} \, , 
\quad \, \, \chi^{a \dot\alpha}  \, \overset{-2 \gamma}{\rightarrow} \,
 \chi^{\alpha \dot{a}} \, .
\end{eqnarray}

\subsection{String and spin chain frames}

In this paper, we compared 
the predictions for the structure
constants obtained using the hexagon program 
with the available weak coupling
data. Thus, it will be convenient to use the spin-chain frame instead 
of the string frame. There is a map between the 
excitations in one frame to the
other and our strategy will be to evaluate the hexagon form factor 
in the string frame using the definitions and crossing rules 
given above and apply the map to the final result. 
Choosing the spin-chain frame parameters conveniently\footnote{More precisely, the set of parameters 
we are considering is $\gamma_i = \dot\gamma_i=\sqrt{i(x^-_i - x^+_i)}$,
$\kappa= i \alpha =1$ and   
the eigenvalue $z$ of the $Z$ marker is $z = 
e^{-\frac{i p}{2}}$ with $p$ the total momentum of the state.},
the map for derivatives $D$, scalars $\Phi$ and fermions $\Psi$ is 
\cite{hexagon,DiagonalII}
\begin{equation}
D_{{\rm{string}}} = D_{{ \rm{spin}}} \, , \quad \Phi_{{\rm{string}}} = 
Z^{\frac{1}{2}}
\Phi_{{ \rm{spin}}} Z^{\frac{1}{2}} \, , \quad  
\Psi_{{\rm{string}}} = 
Z^{\frac{1}{4}}
\Psi_{{ \rm{spin}}} Z^{\frac{1}{4}} \, , \quad  
\end{equation}
where $Z$ is the $Z$ marker. 

As a consequence of the map above, the hexagon form-factors 
computed 
in the string frame can differ
from the ones computed in the spin-chain frame
only by a phase that depends on the momenta of the excitations. 
Using both the rule to pass the $Z$ markers from the right to the left
of an excitation and replacing the $Z$ markers on the left 
of all excitations by their eigenvalues, it is 
possible to derive an expression for this phase, see \cite{hexagon} 
for details. In this work, we are interested in operators with 
fermionic excitations, so we are only going
to give the expression
for the phase in this case. The expression is a generalization
of the one in \cite{hexagon} for scalars and the   
derivation is similar. Consider that the upper edge of the hexagon
has $N_1$ physical excitations
with momenta $p^{(1)}
_i$. In addition, consider that 
the next physical edge moving anticlockwise
has $N_2$ excitations with momenta $p^{(2)}
_j$ and the remaining physical
edge has $N_3$ excitations with momenta $p^{(3)}_k$. 
For this configuration, one has
\begin{eqnarray} \label{conversion}
h(N_1, p^{(1)}_i;N_2, p^{(2)}_j;N_3, p^{(3)}_k) _{{\rm{string}}} =
F_{p^{(1)}} F_{p^{(2)}} 
F_{p^{(3)}} \times \hspace{25mm} 
\label{eq:additionalsigns} \\ \nonumber \\
e^{- \frac{i}{4} [ P^{(1)} (N_1+ N_3-N_2) +
P^{(2)}(N_2+N_1-N_3)+ P^{(3)}(N_3+N_2-N_1)]}  \, \, 
h(N_1, p^{(1)}_i;N_2, p^{(2)}_j;N_3, p^{(3)}_k) _{{\rm{spin}}} \,, 
\nonumber
\end{eqnarray}
where $h$ means hexagon form-factor, 
$P^{(i)}$ is the total momentum of the excitations $i$ and 
\begin{equation}
F_{p^{(i)}} = \prod_{j=1}^{N_i} e^{ \frac{i p^{(i)}_j}{2} 
(N_i + \frac{1}{2} -j)} \, . 
\end{equation}

\end{document}